%% file: paper_main.tex
\documentclass[%
reprint,
superscriptaddress,
amsmath,
amssymb,
gensymb,
aps,
pra,
]{revtex4-2}

\usepackage[utf8]{inputenc}
\usepackage[T1]{fontenc}
\usepackage{color}
\usepackage{graphicx}% Include figure files
\usepackage{dcolumn}% Align table columns on decimal point
\usepackage{bm}% bold math
\usepackage{xfrac}
\usepackage{upgreek}

\usepackage{dcolumn}% Align table columns on decimal point
\usepackage{bm}% bold math
\usepackage{braket}
\usepackage{sidecap}
\usepackage {hyperref}
\sidecaptionvpos{figure}{c}

\graphicspath{Figures}

\begin{document}

%\preprint{APS/123-QED}

\title{High performance single-photon sources at telecom wavelength based on broadband hybrid circular Bragg gratings}% Force line breaks with \\

\author{Andrea Barbiero}
%\thanks{These authors contributed equally to this work}
\email{andrea.barbiero@crl.toshiba.co.uk}
\affiliation{Toshiba Europe Limited, 208 Science Park, Milton Road, Cambridge, CB4 0GZ, UK}
\affiliation{Department of Physics and Astronomy, University of Sheffield, Hounsfield Road, Sheffield, S3 7RH UK}
\author{Jan Huwer}
\affiliation{Toshiba Europe Limited, 208 Science Park, Milton Road, Cambridge, CB4 0GZ, UK}
\author{Joanna Skiba-Szymanska}
\affiliation{Toshiba Europe Limited, 208 Science Park, Milton Road, Cambridge, CB4 0GZ, UK}
\author{David J. P. Ellis}
\affiliation{Toshiba Europe Limited, 208 Science Park, Milton Road, Cambridge, CB4 0GZ, UK}
\author{R. Mark Stevenson}
\affiliation{Toshiba Europe Limited, 208 Science Park, Milton Road, Cambridge, CB4 0GZ, UK}
\author{Tina M\"{u}ller}
\affiliation{Toshiba Europe Limited, 208 Science Park, Milton Road, Cambridge, CB4 0GZ, UK}
\author{Ginny Shooter}
\affiliation{Toshiba Europe Limited, 208 Science Park, Milton Road, Cambridge, CB4 0GZ, UK}
\author{Lucy E. Goff}
\affiliation{Cavendish Laboratory, University of Cambridge, Madingley Road, Cambridge, CB3 0HE, United Kingdom}
\author{David A. Ritchie}
\affiliation{Cavendish Laboratory, University of Cambridge, Madingley Road, Cambridge, CB3 0HE, United Kingdom}
\author{Andrew J. Shields}
\affiliation{Toshiba Europe Limited, 208 Science Park, Milton Road, Cambridge, CB4 0GZ, UK}

\date{\today}

\begin{abstract}
\noindent Semiconductor quantum dots embedded in hybrid circular Bragg gratings are a promising platform for the efficient generation of nonclassical light.
The scalable fabrication of multiple devices with similar performance is highly desirable for their practical use as sources of single and entangled photons, while the ability to operate at telecom wavelength is essential for their integration with the existing fiber infrastructure.
In this work we combine the promising properties of broadband hybrid circular Bragg gratings with a membrane-transfer process performed on 3" wafer scale. We develop and study single-photon sources based on InAs/GaAs quantum dots emitting in the telecom O-band, demonstrating bright single-photon emission with Purcell factor > 5 and count rates up to 10 MHz.
Furthermore, we address the question of reproducibility by benchmarking the performance of 10 devices covering a wide spectral range of 50 nm within the O-band.
\end{abstract}

\maketitle

%\tableofcontents

%\include{Introduction/Introduction}
\input{Section_1/Section_1}

\input{Section_2/Section_2}

\input{Section_3/Section_3}
\input{Section_4/Section_4}

%\input{Section_5/Section_5}
%\appendix
%\input{Appendix/appendix.tex}

\begin{acknowledgments}
\noindent The authors acknowledge funding from the Ministry of Internal Affairs and Communications, Japan, via the project of ICT priority technology (JPMI00316) ‘Research and Development for Building a Global Quantum Cryptography Communication Network’.
A.B. thanks Aleksander Tartakovskii for the academic supervision.
%\smallskip
R.M.S, A.J.S and D.A.R. guided and supervised the project. L.E.G. engineered the epitaxial process and grew the sample. A.B. optimized the device design using FEM simulations.  A.B., J.S.-S. and D.J.P.E. fabricated the devices. A.B. carried out spectroscopic measurements and analysed the data with support from J.H., R.M.S., T.M. and G.S. A.B. wrote the manuscript. All authors discussed the results and commented on the manuscript.
%\smallskip

%The authors declare that they have no competing financial interests.
%
%\noindent The authors declare that they have no competing financial interests.
\end{acknowledgments}

% The \nocite command causes all entries in a bibliography to be printed out
% whether or not they are actually referenced in the text. This is appropriate
% for the sample file to show the different styles of references, but authors
% most likely will not want to use it.
%\nocite{*}

%\bibliographystyle{naturemag} % Use for unsorted references
%\bibliographystyle{plainnat} % use this to have URLs listed in References

\bibliography{paper_bibliography}

\end{document}

%% file: Section_1/Section_1.tex
\section{Introduction}
\noindent Emerging photonic quantum technologies, such as quantum key distribution \cite{Gisin.2002, H.S.Singh.2014}, quantum communication networks \cite{Kimble.2008, Xu.2020} and quantum computing \cite{Kok.2007,Slussarenko.2019}, require the efficient generation of photons with high purity and indistinguishability. The last few years have seen an encouraging progress in the development of bright single-photon sources based on the combination of semiconductor quantum dots (QDs) \cite{Arakawa.2020} with vertically emitting photonic structures such as nanowires \cite{Claudon.2010, Haffouz.2018, Kotal.2021}, cavities \cite{Moreau.2001b, Santori.2002, Somaschi.2016, Thomas.2021b, Tomm.2021} or microlenses \cite{Gschrey.2015, Sartison.2017, Chen.2018}.

Hybrid circular Bragg gratings (CBGs) \cite{Yao.2018} have recently emerged as a promising alternative to those platforms thanks to the presence of a backside gold mirror, which eliminates the loss of photons into the substrate as well as the issue of a fragile free-standing membrane that affects the original suspended version \cite{Davanco.2011, Sapienza.2015, Kolatschek.2019}.
The first experiments with hybrid CBGs have shown the ability to provide a moderate Purcell enhancement of both the exciton and the biexciton transitions as well as high extraction efficiency in a wide range of wavelengths \cite{Liu.2019, Wang.2019}, leading to count rates > 10 MHz \cite{Wang.2019c, You.2021}. However, they focused on the spectral region between 780 nm and 880 nm, which prevents their integration with the standard optical fiber infrastructure due to strong attenuation (> 1 dB/km) at those wavelengths.
Despite the publication of various design studies \cite{Rickert.2019,Blokhin.2021, Barbiero.2022, Bremer.2022} and the demonstration of a device operating in the O-band \cite{Kolatschek.2021}, similar results at telecom wavelength have not been achieved yet and the degree of reproducibility remains an open question.
\begin{figure}[h]
\includegraphics[width=0.48\textwidth]{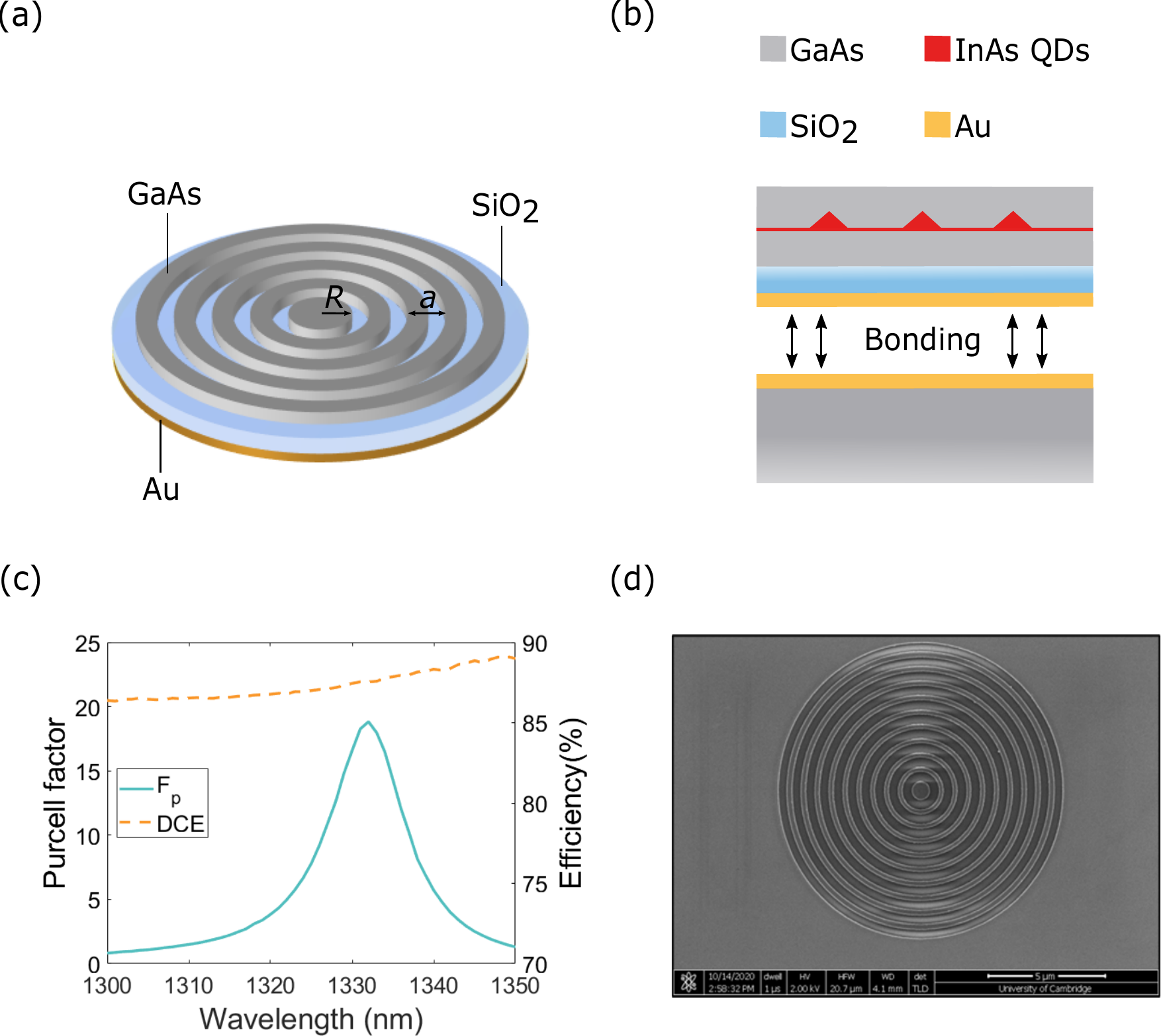}
\caption{\textbf{Structure and simulated performance of a broadband hybrid CBG.} (a) 3D illustration of the device, which consists of a GaAs disk and circular grating, an insulating layer of SiO\textsubscript{2} and a backside Au mirror. $R$ and $a$ indicate the radius of the central disk and the periodicity of the grating, respectively. (b) Schematic cross-section showing the layer stack of the bonded sample after lapping, polishing and selective wet etching. (c) Simulated performance of a device operating in the telecom O-band, with Purcell factor (teal solid line) and dipole collection efficiency (DCE) in NA = 0.65 (orange dashed line) as a function of wavelength. (d) SEM top view of an exemplary device fabricated on the bonded GaAs slab.
\label{Fig1} }
\end{figure}
\begin{figure*}
\includegraphics[width=0.69\textwidth]{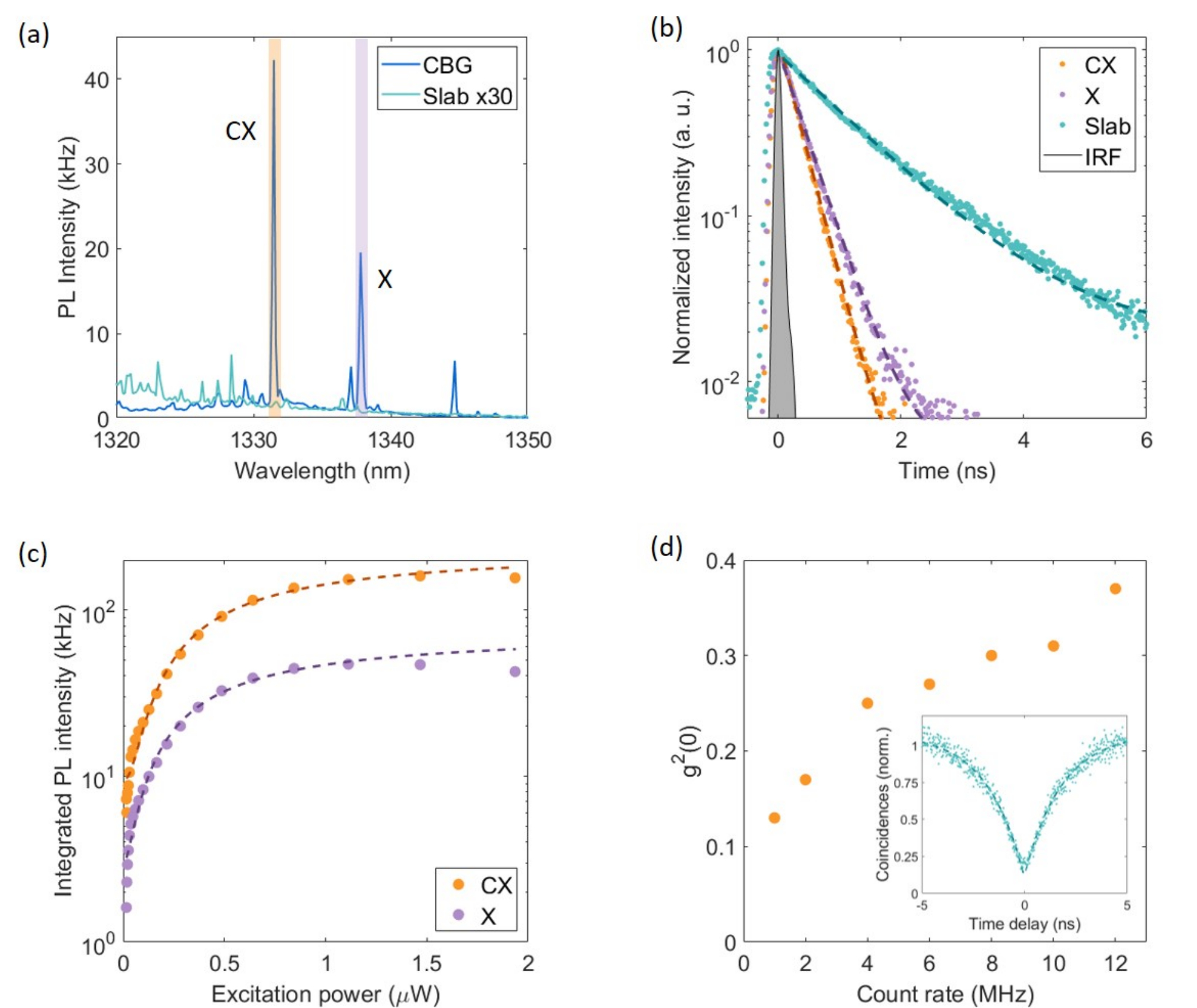}
\caption{\textbf{Characterization of a broadband hybrid CBG.} (a) Micro-PL signal from an InAs QD embedded in the CBG (blue) and from QDs in an unetched region of the GaAs membrane outside the device (teal) under above-band CW laser excitation. The signal acquired on the bare slab is multiplied by a factor of 30. The charged (CX) and neutral (X) excitonic transitions investigated are marked in orange and violet, respectively. (b) Decay traces of the CX transition (orange), of the X transition (violet) and of an exemplary reference line from the unetched GaAs slab (teal). The grey area indicates the instrument response function (IRF). The dashed lines represent the exponential fit of the decay traces. (c) Background corrected saturation curves of the CX (orange) and X (violet) transitions. The data points indicate the integrated intensity of the CX and X spectral lines measured with the spectrometer. (d) Extracted values of the $g\textsuperscript{2}(0)$ for the CX transition (orange) as a function of the combined photon rate measured using SNSPDs with $\sim$ 60\% efficiency. The inset shows an example of second order photon correlation (teal) measured at a rate of 1 MHz.
\label{Fig2} }
\end{figure*}
\\In this framework, a recently proposed broadband version of the hybrid CBG structure \cite{Rickert.2019} has revealed potential advantages towards the reproducible fabrication of multiple devices with similar optical properties. In fact, numerical simulations suggested that this design variation should maintain excellent performance despite its less demanding fine features and guarantee an improved robustness against common imperfections, such as tilted side walls or spatial displacement of the QD emitter.

Here, we employ a membrane transfer process on the wafer scale to fabricate broadband hybrid CBGs and harness their promising properties for the efficient generation of quantum light at telecom wavelength.
By investigating the photoluminescence (PL) signal from self-assembled InAs/GaAs QDs embedded in these devices, we report bright single-photon emission in the telecom O-band with Purcell factor > 5 and count rates up to 10 MHz under above-band excitation.
We also show the simultaneous enhancement of multiple transitions, with Purcell factor > 4 over a 7 nm bandwidth.
Moreover, we demonstrate that the non-classical nature of the emitted light is preserved even after the propagation through up to 50 km of standard optical fiber.
Finally, we address the question of reproducibility by characterizing 10 devices with similar design parameters.

%% file: Section_2/Section_2.tex
% model section
\section{Device design and fabrication}
\noindent A hybrid CBG consists of concentric trenches etched with periodicity $a$ in a semiconductor membrane around a central disk of radius $R$ (Figure \ref{Fig1}a). The device incorporates QDs vertically centered in the slab and includes also a backside gold mirror and an insulating oxide layer.
Our devices are based on epitaxially grown self-assembled InAs QDs embedded in a 240 nm thick GaAs membrane, which sits on a 200 nm thick Al\textsubscript{0.8}Ga\textsubscript{0.2}As sacrificial layer. These layers have been deposited on an undoped GaAs (100) substrate.

In order to facilitate the scalable production of broadband CBGs, we employed a membrane-transfer technique based of thermocompression wafer bonding followed by mechanical lapping and wet chemical etching.
In this process, the GaAs slab is covered with a film of SiO\textsubscript{2} using plasma enhanced chemical vapour deposition (PECVD) and approximately 200 nm of Au is subsequently evaporated on top of both the SiO\textsubscript{2} layer and a GaAs (100) carrier substrate. Next, the two metallic surfaces are brought into atomic contact, applying force and heat simultaneously in order to generate a continuous bond between the interfaces by solid-state diffusion \cite{Reinert.2015}.
After the bonding, the majority of the original GaAs substrate is reduced to an approximate thickness of 30 $\mu$m by commercial mechanical lapping and polishing. The leftover material is then etched in a solution of citric acid (C\textsubscript{6}H\textsubscript{8}O\textsubscript{7}) and hydrogen peroxide (H\textsubscript{2}O\textsubscript{2}) with a 5:1 ratio, which is highly selective \cite{DeSalvo.1992, Moon.1998} and stops at the  interface with the Al\textsubscript{0.8}Ga\textsubscript{0.2}As sacrificial layer.
A final dip in diluted HCl dissolves the Al\textsubscript{0.8}Ga\textsubscript{0.2}As, yielding the layer stack shown in Figure \ref{Fig1}b.
This whole process is performed on 3" wafer scale, producing a large amount of material ready for the fabrication of CBGs. Moreover, combining mechanical lapping with a gentle and controlled wet etching guarantees low surface roughness and minimizes the number of growth and processing steps required to integrate the backside gold mirror \cite{Fischbach.2017,Schmidt.2020,Srocka.2020}.

The thickness of the GaAs and SiO\textsubscript{2} layers and the design parameters of the broadband CBGs are optimized using 3D FEM numerical simulations. As shown in Figure \ref{Fig1}c, it is possible to achieve a collection efficiency in NA = 0.65 above 85\% across a bandwidth > 50 nm, in addition to a cavity mode centred around 1330 nm with Purcell factor $F\textsubscript{p} \sim$ 19 and $FWHM$ $\sim$ 12 nm.
It is worth noting that the resonance wavelength of these devices can be easily adjusted by scaling $R$ and $a$ with minimal impact on the collection efficiency \cite{Liu.2019,Wang.2019,Rickert.2019, Barbiero.2022}.
\\The CBGs are fabricated on the bonded GaAs slab using electron-beam lithography and a chlorine-based dry etch process (Figure \ref{Fig1}d).

%% file: Section_3/Section_3.tex
\section{Results}
\noindent For optical characterization, the devices are mounted inside a closed-cycle cryostat at a temperature of 5K and the QDs are excited by either a CW or a pulsed laser diode at a wavelength of 785 nm. The emitted photons are collected with a fiber-coupled confocal microscope mounting a NA=0.5 objective lens and sent to a spectrometer equipped with an InGaAs photodiode array for analysis.
%%
%\begin{figure}
%\includegraphics[width=0.4\textwidth]{Figures/hero_dot_lifetimes_2panels_vertical.jpg}
%\caption{(a) Micro-PL signal from an InAs QD embedded in a CBG (blue) and from an unetched region of the GaAs slab outside the device (teal) under above-band CW laser excitation. The signal acquired on the bare slab is multiplied by a factor of 30. The charged (CX) and neutral (X) excitonic transitions investigated are marked in orange and violet, respectively. (b) Decay traces of the CX transition (orange), of the X transition (violet) and of an exemplary reference line from the unetched GaAs slab (teal). The grey area indicates the instrument response function (IRF). The dashed lines represent the exponential fit of the decay traces.
%\label{Fig2} }
%\end{figure}
%%

Figure \ref{Fig2}a shows the photoluminescence emission of a single InAs QD in a broadband hybrid CBG under CW laser excitation. The spectrum reveals a clear enhancement of the PL intensity by approximately two orders of magnitude compared to quantum dots in the bare slab. Such an enhancement originates from both the efficient out-of-plane coupling guaranteed by the grating and the resonant coupling of the QD emission to a cavity mode.
The relatively low density of spectral lines in our devices allowed us to identify neutral and charged exciton transitions using polarization-resolved spectroscopy.

To probe the overlap between the QD transitions and the cavity mode and to quantify the Purcell enhancement provided by the CBG, we studied the temporal evolution of the carrier populations. The light collected from the sample was spectrally filtered with a diffraction grating (62\% efficiency, $FWHM$ = 0.58 nm) and detected using superconducting nanowires single photon detectors (SNSPDs).
Initially, we created a baseline by measuring the radiatiative lifetimes of 18 transitions from 5 different unetched areas of the GaAs slab, obtaining an average lifetime of $\tau\textsuperscript{slab}$ = 1.594 ns and a standard deviation $\sigma\textsuperscript{slab}$ = 254 ps.
In Figure \ref{Fig2}b we compare a typical time-resolved luminescence trace measured on the slab with the one of the charged exciton (CX) transition at $\lambda$ = 1331 nm, which decays with a time constant of 289 ps. As a result, we estimate a Purcell factor of $\sim$ 5.5.
A similar decay time of 358 ps is observed for the neutrally charged (X) transition at $\lambda$ = 1337 nm, demonstrating that the device can enhance the radiative decay rate by a factor > 4 over a wavelength range of 6 nm.

The power dependence in Figure \ref{Fig2}c, obtained by integrating the PL intensity recorded with the spectrometer, indicates that the QD emission saturates for a CW excitation power $P\textsubscript{sat}$ of approximately 1 $\mu$W. The corresponding SNSPD count rate recorded at saturation exceeds 10 MHz for the CX transition.
Furthermore, we measured a remarkable rate of 2 MHz under pulsed excitation at 80 MHz, which places this device among the brightest QD-based single photon sources at telecom wavelength reported in literature \cite{Kim.2016, Haffouz.2018, Srocka.2020, Yang.2020, Morrison.2021, Kolatschek.2021, Zeuner.2021, DaLio.2022b}.
The efficiency of the source system at the output of the spectral filter can be consequently estimated by removing the 2.3 dB loss of the detection system and the 0.13 dB loss introduced by the optics: a CX photon is produced by the source system in 4.3\% of the clock cycles, with a photon rate of 3.4 MHz.
%Considering the efficiency of the SNSPDs ($\sim$ 60\%) and the losses introduced by the optics (-0.13 dB), this corresponds to a 4.3\% efficiency of the source system (i.e. hybrid CBG, fiber-coupled confocal microscope and spectral filter).

To evaluate the single-photon purity of the emission, we sent the spectrally filtered light to a fiber-based Hanbury Brown and Twiss interferometer and measured the second-order correlation function $g^{(2)}(\tau)$ for increasing CW laser power. Figure \ref{Fig2}d summarizes the results of this study, showing the $g^{(2)}(0)$ as a function of the combined photon rate recorded on the detectors. The occurrence of multi-photon events is found to increase with higher excitation power, due to saturation of the transition examined and increasing background emission. However, the results confirm that the condition $g^{(2)}(0)$ < 0.5 is fulfilled for count rates > 10 MHz.
%
%\begin{figure} [h]
%\includegraphics[width=0.4\textwidth]{Figures/g2_vs_count_rate.jpg}
%\caption{Extracted values of $g\textsuperscript{2}(0)$ for the CX transition as a function of the photon rate measured using SNSPDs with $\sim$ 60\% efficiency. The inset shows an example of second order photon correlation measured at a rate of 1 MHz.
%\label{Fig3} }
%\end{figure}
%
%
%\begin{figure} [h]
%\includegraphics[width=0.4\textwidth]{Figures/hero_dot_PwDep_and_count_rate.jpg}
%\caption{(a) Background corrected saturation curves of the CX and X transitions. The data points indicate the integrated intensity of the CX and X spectral lines measured with the spectrometer. (b) Extracted values of $g\textsuperscript{2}(0)$ for the CX transition as a function of the combined photon rate measured using SNSPDs with $\sim$ 60\% efficiency. The inset shows an example of second order photon correlation measured at a rate of 1 MHz.
%\label{Fig3} }
%\end{figure}
%
In the future, the $g^{(2)}(0)$ could be improved by adopting resonant excitation schemes, which have been shown to minimise multiphoton events and lead to essentially background-free single photon emission \cite{Wang.2019, Wang.2019c}.

\begin{figure} [h]
\includegraphics[width=0.42\textwidth]{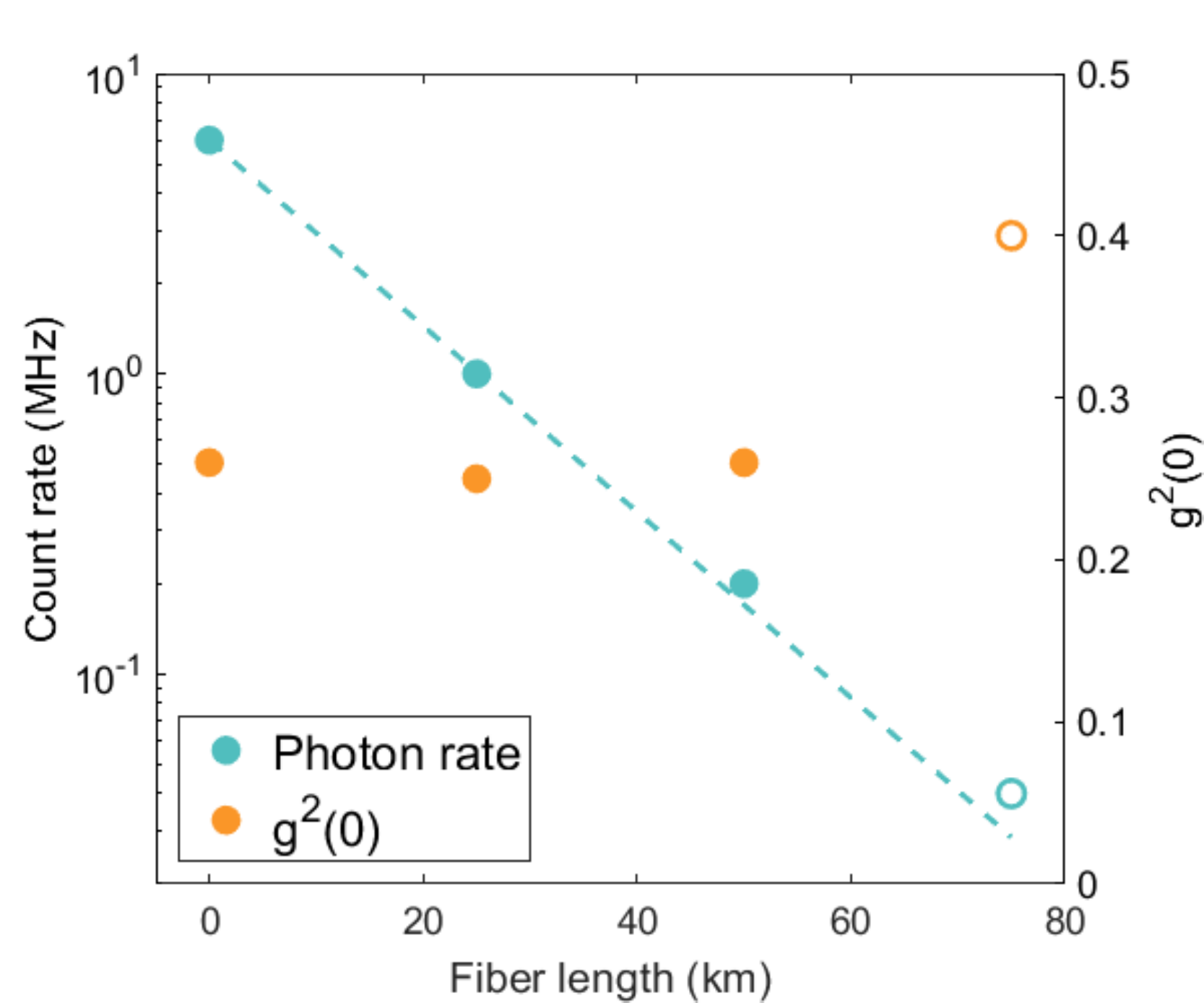}
\caption{\textbf{Second-order correlation measurements after propagation through SM fiber.} Photon rates (teal) and corresponding values of the $g\textsuperscript{2}(0)$ (orange) as a function of the fiber length between the source and the SNSPDs measured for the CX transition. The full circles correspond to data taken at 0.75$P\textsubscript{sat}$ while the hollow ones represent data taken at $P\textsubscript{sat}$. The attenuation value of 0.31 dB/km is extracted by fitting the data points measured at 0.75$P\textsubscript{sat}$ (teal dashed line).
\label{Fig4} }
\end{figure}

To underline that the improved brightness provided by the broadband hybrid CBG combined with the telecom wavelength range of the source can be advantageous for long-haul transmission of quantum light and operation over a fiber network, we repeated the second-order correlation measurements after adding a variable length of commercial single-mode (SM) telecom fiber between the source and the detectors.
In Figure \ref{Fig4} we report the combined count rate recorded on the SNSPDs and the corresponding value of the $g^{(2)}(0)$ as a function of the fiber length when the device is pumped with a CW laser power of 0.75$P\textsubscript{sat}$. While the signal intensity drops exponentially with distance because of the attenuation $\alpha$ = 0.31 dB/km, the $g^{(2)}(0)$ remains almost constant around 0.25 for a fiber length up to 50 km. This demonstrates that the non-classical nature of the quantum dot emission is preserved over long distances, with a remarkable photon rate of 200 kHz recorded after propagation through 50 km of fiber.
When moving to 75 km, the pump power was raised to $P\textsubscript{sat}$ to probe the maximum signal intensity achievable at long distances. As a result, we measured a photon rate of 40 kHz and an increased $g^{(2)}(0)= 0.40$ due to saturation of the QD transition and increasing background contribution.
\begin{figure*}
\includegraphics[width=1\textwidth]{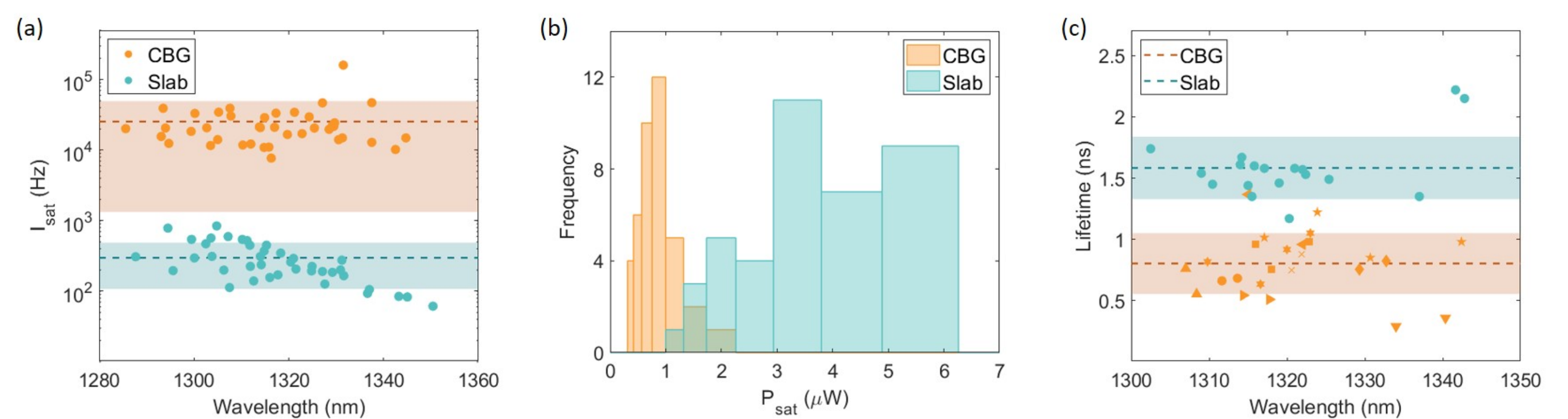}
\caption{\textbf{Performance of 10  broadband hybrid CBGs operating in the telecom O-band.} (a) Integrated PL intensity at saturation $I\textsubscript{sat}$ measured with the spectrometer for 40 transitions on 10 different CBGs (orange) and 40 transitions on the unetched GaAs membrane (teal). The dashed lines and shaded areas represent the mean values and corresponding standard deviation, respectively. (b) Corresponding CW laser power $P\textsubscript{sat}$ required to reach saturation of the QD transitions. (c) Radiative lifetimes of 25 transitions measured on 10 different CBGs (orange) and 18 transitions measured on the unetched slab (teal). The dashed lines and shaded areas represent the mean values and corresponding standard deviation, respectively. Each shape of the orange markers indicates data taken on a specific device.
\label{Fig5} }
\end{figure*}

As mentioned above, the fabrication of multiple devices with similar performance is highly desirable for the large-scale diffusion of QD-based single-photon sources. Therefore, we analysed the performance of 10 different broadband CBGs with design parameters selected to shift the resonance wavelength over a wide spectral range of 50 nm within the telecom O-band.
In Figure \ref{Fig5}a we compare the integrated PL intensity at saturation $I\textsubscript{sat}$ for 40 transitions measured on those CBGs (4 data points per device) and 40 transitions measured on various positions of the unetched slab.
The average PL intensity at saturation recorded on the bare slab is  $I\textsubscript{sat}\textsuperscript{slab}$ = 297 cts/s with a standard deviation of 190 cts/s. As for the CBGs, we observe an average intensity $I\textsubscript{sat}\textsuperscript{CBG}$ = 24129 cts/s (standard deviation: 22129 cts/s), corresponding to an average enhancement factor of almost two orders of magnitude with no dependence on the operational wavelength. The Purcell-enhanced CX transition shown in Figure \ref{Fig2}a exhibits the brightest PL signal ($I\textsubscript{sat} \sim$ 160470 cts/s), corresponding to a remarkable enhancement factor of 540 compared to QDs in the bare slab.
It is also important to underline that some of the transitions included in Figure \ref{Fig5}a are not Purcell enhanced: nevertheless, they still benefit from the improved extraction efficiency over a broad spectral range guaranteed by the CBGs.
For the sake of rigor, we observe that the thickness of the oxide layer has been optimized taking into account the presence of the circular grating, resulting in an imperfect distance from the backside Au mirror for QDs in the bare slab, while the reduced photon rate compared to the SNSPDs measurements reported above is caused by additional losses from the spectrometer.

Furthermore, we notice that QDs coupled to a CBG require a lower pump power $P\textsubscript{sat}$ to reach saturation (Figure \ref{Fig5}b). We attribute this effect to the presence of the circular grating, which helps to focus the incoming light towards the central disk for a more efficient laser excitation that may reduce the charge noise around the QDs.

Finally, Figure \ref{Fig5}c summarizes the results of time-resolved measurements over multiple devices.
As explained previously, the teal data points indicate the radiative lifetimes of 18 transitions from 5 different locations on the unetched GaAs slab. The corresponding average lifetime of $\tau\textsuperscript{slab}$ = 1.594 ns and standard deviation $\sigma\textsuperscript{slab}$ = 254 ps are represented by the teal dashed line and shaded area, respectively.
As for the 10 CBGs, we selected two to four transitions on each device in the proximity of the cavity mode. The orange data points indicate the corresponding radiative lifetimes, which in most of the cases are clearly faster than the ones measured on the bare slab.
Considering the mean value of $\tau\textsuperscript{CBG}$ = 802 ps and standard deviation $\sigma\textsuperscript{CBG}$ = 248 ps, we estimate an average Purcell factor $\sim$ 2 for the 25 spectral lines examined with no dependence on the operational wavelength. Furthermore, we notice that 8 out of 10 sources display at least one transition with $F\textsubscript{p}$ > 2.
\\These results confirm that the promising properties of broadband CBGs can facilitate the production of multiple devices with similar performance even with a non-deterministic fabrication method. 

%% file: Section_4/Section_4.tex
\section{Conclusions}
%
%\begin{figure*}
%\includegraphics[width=0.7\textwidth]{Figures/polar_plots_fig_alloy_v3.pdf}
%\caption{Polar diagrams showing the square of the optical transition dipole matrix element calculated for GaAs/AlAs (top panel), GaAs/Al\textsubscript{0.15}Ga\textsubscript{0.85}As (middle panel) and InAs/GaAs (bottom panel) QDs assuming different values of the miscut angle $\alpha$. X\textsubscript{LOW} (X\textsubscript{HIGH}) represents the lowest (highest) lying bright exciton state in energy. When two exciton bright states are degenerate, they are labelled X\textsubscript{B}.
%\label{Fig3} }
%\end{figure*}
%%
%
\noindent The efficient generation of quantum light at telecom wavelength is crucial for the success of many quantum communication protocols and the development of fiber-based quantum networks. Moreover, the large-scale uptake of those technologies requires high-yield fabrication of multiple sources with reproducible performance.

In this work, we produced and characterized multiple quantum light sources in the telecom O-band based on self-assembled InAs/GaAs QDs coupled to broadband hybrid circular Bragg gratings.
We introduced a membrane-transfer technique performed on 3" wafer scale, which is an important step towards the large-scale fabrication of those devices and we took full advantage of the broadband nature of the CBGs to obtain multiple sources with reproducible properties even with a non-deterministic lithographic process.
We demonstrated that these devices can provide Purcell factors > 4 over a broad wavelength range of 6 nm, leading to the simultaneous enhancement of multiple transitions and facilitating the spectral overlap with the cavity mode. In addition, we proved that an individual transition can be enhanced by a factor > 5 and provide photon rates exceeding 10 MHz while preserving the single-photon emission properties of the source.
We also showed that the non-classical nature of the emitted light is preserved over long distances, with a remarkable photon rate of 200 kHz recorded after propagation through 50 km of standard SM optical fiber.
Finally, we addressed the question of reproducibility and demonstrated that the optical and quantum properties are replicated for QDs embedded into broadband hybrid CBGs with similar design parameters.

Combining high brightness, Purcell enhancement and reproducibility with emission in the O-band is a significant step towards the scalable production of high performance single-photon sources for quantum network applications. Although we demonstrated that broadband hybrid CBGs operating at telecom wavelength have already the potential to be versatile devices for long-distance quantum communication, future experiments may include applying resonant excitation techniques to reveal the real quality of these sources in terms of purity, indistinguishability and efficiency.
Moreover, the spatial and spectral coupling between the QD emission and the cavity mode could be further optimized by employing in-situ electron beam lithography.

%% file: paper_main.bbl
%apsrev4-2.bst 2019-01-14 (MD) hand-edited version of apsrev4-1.bst
%Control: key (0)
%Control: author (8) initials jnrlst
%Control: editor formatted (1) identically to author
%Control: production of article title (0) allowed
%Control: page (0) single
%Control: year (1) truncated
%Control: production of eprint (0) enabled
\begin{thebibliography}{42}%
\makeatletter
\providecommand \@ifxundefined [1]{%
 \@ifx{#1\undefined}
}%
\providecommand \@ifnum [1]{%
 \ifnum #1\expandafter \@firstoftwo
 \else \expandafter \@secondoftwo
 \fi
}%
\providecommand \@ifx [1]{%
 \ifx #1\expandafter \@firstoftwo
 \else \expandafter \@secondoftwo
 \fi
}%
\providecommand \natexlab [1]{#1}%
\providecommand \enquote  [1]{``#1''}%
\providecommand \bibnamefont  [1]{#1}%
\providecommand \bibfnamefont [1]{#1}%
\providecommand \citenamefont [1]{#1}%
\providecommand \href@noop [0]{\@secondoftwo}%
\providecommand \href [0]{\begingroup \@sanitize@url \@href}%
\providecommand \@href[1]{\@@startlink{#1}\@@href}%
\providecommand \@@href[1]{\endgroup#1\@@endlink}%
\providecommand \@sanitize@url [0]{\catcode `\\12\catcode `\$12\catcode
  `\&12\catcode `\#12\catcode `\^12\catcode `\_12\catcode `\%12\relax}%
\providecommand \@@startlink[1]{}%
\providecommand \@@endlink[0]{}%
\providecommand \url  [0]{\begingroup\@sanitize@url \@url }%
\providecommand \@url [1]{\endgroup\@href {#1}{\urlprefix }}%
\providecommand \urlprefix  [0]{URL }%
\providecommand \Eprint [0]{\href }%
\providecommand \doibase [0]{https://doi.org/}%
\providecommand \selectlanguage [0]{\@gobble}%
\providecommand \bibinfo  [0]{\@secondoftwo}%
\providecommand \bibfield  [0]{\@secondoftwo}%
\providecommand \translation [1]{[#1]}%
\providecommand \BibitemOpen [0]{}%
\providecommand \bibitemStop [0]{}%
\providecommand \bibitemNoStop [0]{.\EOS\space}%
\providecommand \EOS [0]{\spacefactor3000\relax}%
\providecommand \BibitemShut  [1]{\csname bibitem#1\endcsname}%
\let\auto@bib@innerbib\@empty
%</preamble>
\bibitem [{\citenamefont {Gisin}\ \emph {et~al.}(2002)\citenamefont {Gisin},
  \citenamefont {Ribordy}, \citenamefont {Tittel},\ and\ \citenamefont
  {Zbinden}}]{Gisin.2002}%
  \BibitemOpen
  \bibfield  {author} {\bibinfo {author} {\bibfnamefont {N.}~\bibnamefont
  {Gisin}}, \bibinfo {author} {\bibfnamefont {G.}~\bibnamefont {Ribordy}},
  \bibinfo {author} {\bibfnamefont {W.}~\bibnamefont {Tittel}},\ and\ \bibinfo
  {author} {\bibfnamefont {H.}~\bibnamefont {Zbinden}},\ }\bibfield  {title}
  {\bibinfo {title} {{Quantum cryptography}},\ }\href
  {https://doi.org/10.1103/RevModPhys.74.145} {\bibfield  {journal} {\bibinfo
  {journal} {{Reviews of Modern Physics}}\ }\textbf {\bibinfo {volume} {74}},\
  \bibinfo {pages} {145} (\bibinfo {year} {2002})}\BibitemShut {NoStop}%
\bibitem [{\citenamefont {{H. S. Singh}}\ \emph {et~al.}(2014)\citenamefont
  {{H. S. Singh}}, \citenamefont {{Dikshi Gupta}},\ and\ \citenamefont {{A. K.
  Singh}}}]{H.S.Singh.2014}%
  \BibitemOpen
  \bibfield  {author} {\bibinfo {author} {\bibnamefont {{H. S. Singh}}},
  \bibinfo {author} {\bibnamefont {{Dikshi Gupta}}},\ and\ \bibinfo {author}
  {\bibnamefont {{A. K. Singh}}},\ }\bibfield  {title} {\bibinfo {title}
  {{Quantum Key Distribution Protocols: A Review}},\ }\href@noop {} {\bibfield
  {journal} {\bibinfo  {journal} {{IOSR Journal of Computer Engineering}}\
  }\textbf {\bibinfo {volume} {16}},\ \bibinfo {pages} {01} (\bibinfo {year}
  {2014})}\BibitemShut {NoStop}%
\bibitem [{\citenamefont {Kimble}(2008)}]{Kimble.2008}%
  \BibitemOpen
  \bibfield  {author} {\bibinfo {author} {\bibfnamefont {H.~J.}\ \bibnamefont
  {Kimble}},\ }\bibfield  {title} {\bibinfo {title} {{The quantum internet}},\
  }\href {https://doi.org/10.1038/nature07127} {\bibfield  {journal} {\bibinfo
  {journal} {{Nature}}\ }\textbf {\bibinfo {volume} {453}},\ \bibinfo {pages}
  {1023} (\bibinfo {year} {2008})}\BibitemShut {NoStop}%
\bibitem [{\citenamefont {Xu}\ \emph {et~al.}(2020)\citenamefont {Xu},
  \citenamefont {Ma}, \citenamefont {Zhang}, \citenamefont {Lo},\ and\
  \citenamefont {Pan}}]{Xu.2020}%
  \BibitemOpen
  \bibfield  {author} {\bibinfo {author} {\bibfnamefont {F.}~\bibnamefont
  {Xu}}, \bibinfo {author} {\bibfnamefont {X.}~\bibnamefont {Ma}}, \bibinfo
  {author} {\bibfnamefont {Q.}~\bibnamefont {Zhang}}, \bibinfo {author}
  {\bibfnamefont {H.-K.}\ \bibnamefont {Lo}},\ and\ \bibinfo {author}
  {\bibfnamefont {J.-W.}\ \bibnamefont {Pan}},\ }\bibfield  {title} {\bibinfo
  {title} {{Secure quantum key distribution with realistic devices}},\ }\href
  {https://doi.org/10.1103/RevModPhys.92.025002} {\bibfield  {journal}
  {\bibinfo  {journal} {{Reviews of Modern Physics}}\ }\textbf {\bibinfo
  {volume} {92}},\ \bibinfo {pages} {131} (\bibinfo {year} {2020})}\BibitemShut
  {NoStop}%
\bibitem [{\citenamefont {Kok}\ \emph {et~al.}(2007)\citenamefont {Kok},
  \citenamefont {Munro}, \citenamefont {Nemoto}, \citenamefont {Ralph},
  \citenamefont {Dowling},\ and\ \citenamefont {Milburn}}]{Kok.2007}%
  \BibitemOpen
  \bibfield  {author} {\bibinfo {author} {\bibfnamefont {P.}~\bibnamefont
  {Kok}}, \bibinfo {author} {\bibfnamefont {W.~J.}\ \bibnamefont {Munro}},
  \bibinfo {author} {\bibfnamefont {K.}~\bibnamefont {Nemoto}}, \bibinfo
  {author} {\bibfnamefont {T.~C.}\ \bibnamefont {Ralph}}, \bibinfo {author}
  {\bibfnamefont {J.~P.}\ \bibnamefont {Dowling}},\ and\ \bibinfo {author}
  {\bibfnamefont {G.~J.}\ \bibnamefont {Milburn}},\ }\bibfield  {title}
  {\bibinfo {title} {{Linear optical quantum computing with photonic qubits}},\
  }\href {https://doi.org/10.1103/RevModPhys.79.135} {\bibfield  {journal}
  {\bibinfo  {journal} {{Physical review. A, Atomic, molecular, and optical
  physics}}\ }\textbf {\bibinfo {volume} {79}},\ \bibinfo {pages} {135}
  (\bibinfo {year} {2007})}\BibitemShut {NoStop}%
\bibitem [{\citenamefont {Slussarenko}\ and\ \citenamefont
  {Pryde}(2019)}]{Slussarenko.2019}%
  \BibitemOpen
  \bibfield  {author} {\bibinfo {author} {\bibfnamefont {S.}~\bibnamefont
  {Slussarenko}}\ and\ \bibinfo {author} {\bibfnamefont {G.~J.}\ \bibnamefont
  {Pryde}},\ }\bibfield  {title} {\bibinfo {title} {{Photonic quantum
  information processing: A concise review}},\ }\href
  {https://doi.org/10.1063/1.5115814} {\bibfield  {journal} {\bibinfo
  {journal} {{Applied Physics Reviews}}\ }\textbf {\bibinfo {volume} {6}},\
  \bibinfo {pages} {041303} (\bibinfo {year} {2019})}\BibitemShut {NoStop}%
\bibitem [{\citenamefont {Arakawa}\ and\ \citenamefont
  {Holmes}(2020)}]{Arakawa.2020}%
  \BibitemOpen
  \bibfield  {author} {\bibinfo {author} {\bibfnamefont {Y.}~\bibnamefont
  {Arakawa}}\ and\ \bibinfo {author} {\bibfnamefont {M.~J.}\ \bibnamefont
  {Holmes}},\ }\bibfield  {title} {\bibinfo {title} {{Progress in quantum-dot
  single photon sources for quantum information technologies: A broad spectrum
  overview}},\ }\href {https://doi.org/10.1063/5.0010193} {\bibfield  {journal}
  {\bibinfo  {journal} {{Applied Physics Reviews}}\ }\textbf {\bibinfo {volume}
  {7}},\ \bibinfo {pages} {021309} (\bibinfo {year} {2020})}\BibitemShut
  {NoStop}%
\bibitem [{\citenamefont {Claudon}\ \emph {et~al.}(2010)\citenamefont
  {Claudon}, \citenamefont {Bleuse}, \citenamefont {Malik}, \citenamefont
  {Bazin}, \citenamefont {Jaffrennou}, \citenamefont {Gregersen}, \citenamefont
  {Sauvan}, \citenamefont {Lalanne},\ and\ \citenamefont
  {G{\'e}rard}}]{Claudon.2010}%
  \BibitemOpen
  \bibfield  {author} {\bibinfo {author} {\bibfnamefont {J.}~\bibnamefont
  {Claudon}}, \bibinfo {author} {\bibfnamefont {J.}~\bibnamefont {Bleuse}},
  \bibinfo {author} {\bibfnamefont {N.~S.}\ \bibnamefont {Malik}}, \bibinfo
  {author} {\bibfnamefont {M.}~\bibnamefont {Bazin}}, \bibinfo {author}
  {\bibfnamefont {P.}~\bibnamefont {Jaffrennou}}, \bibinfo {author}
  {\bibfnamefont {N.}~\bibnamefont {Gregersen}}, \bibinfo {author}
  {\bibfnamefont {C.}~\bibnamefont {Sauvan}}, \bibinfo {author} {\bibfnamefont
  {P.}~\bibnamefont {Lalanne}},\ and\ \bibinfo {author} {\bibfnamefont {J.-M.}\
  \bibnamefont {G{\'e}rard}},\ }\bibfield  {title} {\bibinfo {title} {{A highly
  efficient single-photon source based on a quantum dot in a photonic
  nanowire}},\ }\href {https://doi.org/10.1038/nphoton.2009.287x} {\bibfield
  {journal} {\bibinfo  {journal} {{Nature Photonics}}\ }\textbf {\bibinfo
  {volume} {4}},\ \bibinfo {pages} {174} (\bibinfo {year} {2010})}\BibitemShut
  {NoStop}%
\bibitem [{\citenamefont {Haffouz}\ \emph {et~al.}(2018)\citenamefont
  {Haffouz}, \citenamefont {Zeuner}, \citenamefont {Dalacu}, \citenamefont
  {Poole}, \citenamefont {Lapointe}, \citenamefont {Poitras}, \citenamefont
  {Mnaymneh}, \citenamefont {Wu}, \citenamefont {Couillard}, \citenamefont
  {Korkusinski}, \citenamefont {Sch{\"o}ll}, \citenamefont {J{\"o}ns},
  \citenamefont {Zwiller},\ and\ \citenamefont {Williams}}]{Haffouz.2018}%
  \BibitemOpen
  \bibfield  {author} {\bibinfo {author} {\bibfnamefont {S.}~\bibnamefont
  {Haffouz}}, \bibinfo {author} {\bibfnamefont {K.~D.}\ \bibnamefont {Zeuner}},
  \bibinfo {author} {\bibfnamefont {D.}~\bibnamefont {Dalacu}}, \bibinfo
  {author} {\bibfnamefont {P.~J.}\ \bibnamefont {Poole}}, \bibinfo {author}
  {\bibfnamefont {J.}~\bibnamefont {Lapointe}}, \bibinfo {author}
  {\bibfnamefont {D.}~\bibnamefont {Poitras}}, \bibinfo {author} {\bibfnamefont
  {K.}~\bibnamefont {Mnaymneh}}, \bibinfo {author} {\bibfnamefont
  {X.}~\bibnamefont {Wu}}, \bibinfo {author} {\bibfnamefont {M.}~\bibnamefont
  {Couillard}}, \bibinfo {author} {\bibfnamefont {M.}~\bibnamefont
  {Korkusinski}}, \bibinfo {author} {\bibfnamefont {E.}~\bibnamefont
  {Sch{\"o}ll}}, \bibinfo {author} {\bibfnamefont {K.~D.}\ \bibnamefont
  {J{\"o}ns}}, \bibinfo {author} {\bibfnamefont {V.}~\bibnamefont {Zwiller}},\
  and\ \bibinfo {author} {\bibfnamefont {R.~L.}\ \bibnamefont {Williams}},\
  }\bibfield  {title} {\bibinfo {title} {{Bright Single InAsP Quantum Dots at
  Telecom Wavelengths in Position-Controlled InP Nanowires: The Role of the
  Photonic Waveguide}},\ }\href {https://doi.org/10.1021/acs.nanolett.8b00550}
  {\bibfield  {journal} {\bibinfo  {journal} {{Nano letters}}\ }\textbf
  {\bibinfo {volume} {18}},\ \bibinfo {pages} {3047} (\bibinfo {year}
  {2018})}\BibitemShut {NoStop}%
\bibitem [{\citenamefont {Kotal}\ \emph {et~al.}(2021)\citenamefont {Kotal},
  \citenamefont {Artioli}, \citenamefont {Wang}, \citenamefont {Osterkryger},
  \citenamefont {Finazzer}, \citenamefont {Fons}, \citenamefont {Genuist},
  \citenamefont {Bleuse}, \citenamefont {G{\'e}rard}, \citenamefont
  {Gregersen},\ and\ \citenamefont {Claudon}}]{Kotal.2021}%
  \BibitemOpen
  \bibfield  {author} {\bibinfo {author} {\bibfnamefont {S.}~\bibnamefont
  {Kotal}}, \bibinfo {author} {\bibfnamefont {A.}~\bibnamefont {Artioli}},
  \bibinfo {author} {\bibfnamefont {Y.}~\bibnamefont {Wang}}, \bibinfo {author}
  {\bibfnamefont {A.~D.}\ \bibnamefont {Osterkryger}}, \bibinfo {author}
  {\bibfnamefont {M.}~\bibnamefont {Finazzer}}, \bibinfo {author}
  {\bibfnamefont {R.}~\bibnamefont {Fons}}, \bibinfo {author} {\bibfnamefont
  {Y.}~\bibnamefont {Genuist}}, \bibinfo {author} {\bibfnamefont
  {J.}~\bibnamefont {Bleuse}}, \bibinfo {author} {\bibfnamefont {J.-M.}\
  \bibnamefont {G{\'e}rard}}, \bibinfo {author} {\bibfnamefont
  {N.}~\bibnamefont {Gregersen}},\ and\ \bibinfo {author} {\bibfnamefont
  {J.}~\bibnamefont {Claudon}},\ }\bibfield  {title} {\bibinfo {title} {{A
  nanowire optical nanocavity for broadband enhancement of spontaneous
  emission}},\ }\href {https://doi.org/10.1063/5.0045834} {\bibfield  {journal}
  {\bibinfo  {journal} {{Applied Physics Letters}}\ }\textbf {\bibinfo {volume}
  {118}},\ \bibinfo {pages} {194002} (\bibinfo {year} {2021})}\BibitemShut
  {NoStop}%
\bibitem [{\citenamefont {Moreau}\ \emph {et~al.}(2001)\citenamefont {Moreau},
  \citenamefont {Robert}, \citenamefont {G{\'e}rard}, \citenamefont {Abram},
  \citenamefont {Manin},\ and\ \citenamefont {Thierry-Mieg}}]{Moreau.2001b}%
  \BibitemOpen
  \bibfield  {author} {\bibinfo {author} {\bibfnamefont {E.}~\bibnamefont
  {Moreau}}, \bibinfo {author} {\bibfnamefont {I.}~\bibnamefont {Robert}},
  \bibinfo {author} {\bibfnamefont {J.~M.}\ \bibnamefont {G{\'e}rard}},
  \bibinfo {author} {\bibfnamefont {I.}~\bibnamefont {Abram}}, \bibinfo
  {author} {\bibfnamefont {L.}~\bibnamefont {Manin}},\ and\ \bibinfo {author}
  {\bibfnamefont {V.}~\bibnamefont {Thierry-Mieg}},\ }\bibfield  {title}
  {\bibinfo {title} {{Single-mode solid-state single photon source based on
  isolated quantum dots in pillar microcavities}},\ }\href
  {https://doi.org/10.1063/1.1415346} {\bibfield  {journal} {\bibinfo
  {journal} {{Applied Physics Letters}}\ }\textbf {\bibinfo {volume} {79}},\
  \bibinfo {pages} {2865} (\bibinfo {year} {2001})}\BibitemShut {NoStop}%
\bibitem [{\citenamefont {Santori}\ \emph {et~al.}(2002)\citenamefont
  {Santori}, \citenamefont {Fattal}, \citenamefont {Vu{\v{c}}kovi{\'c}},
  \citenamefont {Solomon},\ and\ \citenamefont {Yamamoto}}]{Santori.2002}%
  \BibitemOpen
  \bibfield  {author} {\bibinfo {author} {\bibfnamefont {C.}~\bibnamefont
  {Santori}}, \bibinfo {author} {\bibfnamefont {D.}~\bibnamefont {Fattal}},
  \bibinfo {author} {\bibfnamefont {J.}~\bibnamefont {Vu{\v{c}}kovi{\'c}}},
  \bibinfo {author} {\bibfnamefont {G.~S.}\ \bibnamefont {Solomon}},\ and\
  \bibinfo {author} {\bibfnamefont {Y.}~\bibnamefont {Yamamoto}},\ }\bibfield
  {title} {\bibinfo {title} {{Indistinguishable photons from a single-photon
  device}},\ }\href {https://doi.org/10.1038/nature01086} {\bibfield  {journal}
  {\bibinfo  {journal} {{Nature}}\ }\textbf {\bibinfo {volume} {419}},\
  \bibinfo {pages} {594} (\bibinfo {year} {2002})}\BibitemShut {NoStop}%
\bibitem [{\citenamefont {Somaschi}\ \emph {et~al.}(2016)\citenamefont
  {Somaschi}, \citenamefont {Giesz}, \citenamefont {de~Santis}, \citenamefont
  {Loredo}, \citenamefont {Almeida}, \citenamefont {Hornecker}, \citenamefont
  {Portalupi}, \citenamefont {Grange}, \citenamefont {Ant{\'o}n}, \citenamefont
  {Demory}, \citenamefont {G{\'o}mez}, \citenamefont {Sagnes}, \citenamefont
  {Lanzillotti-Kimura}, \citenamefont {Lema{\'i}tre}, \citenamefont {Auffeves},
  \citenamefont {White}, \citenamefont {Lanco},\ and\ \citenamefont
  {Senellart}}]{Somaschi.2016}%
  \BibitemOpen
  \bibfield  {author} {\bibinfo {author} {\bibfnamefont {N.}~\bibnamefont
  {Somaschi}}, \bibinfo {author} {\bibfnamefont {V.}~\bibnamefont {Giesz}},
  \bibinfo {author} {\bibfnamefont {L.}~\bibnamefont {de~Santis}}, \bibinfo
  {author} {\bibfnamefont {J.~C.}\ \bibnamefont {Loredo}}, \bibinfo {author}
  {\bibfnamefont {M.~P.}\ \bibnamefont {Almeida}}, \bibinfo {author}
  {\bibfnamefont {G.}~\bibnamefont {Hornecker}}, \bibinfo {author}
  {\bibfnamefont {S.~L.}\ \bibnamefont {Portalupi}}, \bibinfo {author}
  {\bibfnamefont {T.}~\bibnamefont {Grange}}, \bibinfo {author} {\bibfnamefont
  {C.}~\bibnamefont {Ant{\'o}n}}, \bibinfo {author} {\bibfnamefont
  {J.}~\bibnamefont {Demory}}, \bibinfo {author} {\bibfnamefont
  {C.}~\bibnamefont {G{\'o}mez}}, \bibinfo {author} {\bibfnamefont
  {I.}~\bibnamefont {Sagnes}}, \bibinfo {author} {\bibfnamefont {N.~D.}\
  \bibnamefont {Lanzillotti-Kimura}}, \bibinfo {author} {\bibfnamefont
  {A.}~\bibnamefont {Lema{\'i}tre}}, \bibinfo {author} {\bibfnamefont
  {A.}~\bibnamefont {Auffeves}}, \bibinfo {author} {\bibfnamefont {A.~G.}\
  \bibnamefont {White}}, \bibinfo {author} {\bibfnamefont {L.}~\bibnamefont
  {Lanco}},\ and\ \bibinfo {author} {\bibfnamefont {P.}~\bibnamefont
  {Senellart}},\ }\bibfield  {title} {\bibinfo {title} {{Near-optimal
  single-photon sources in the solid state}},\ }\href
  {https://doi.org/10.1038/nphoton.2016.23} {\bibfield  {journal} {\bibinfo
  {journal} {{Nature Photonics}}\ }\textbf {\bibinfo {volume} {10}},\ \bibinfo
  {pages} {340} (\bibinfo {year} {2016})}\BibitemShut {NoStop}%
\bibitem [{\citenamefont {Thomas}\ \emph {et~al.}(2021)\citenamefont {Thomas},
  \citenamefont {Billard}, \citenamefont {Coste}, \citenamefont {Wein},
  \citenamefont {Priya}, \citenamefont {Ollivier}, \citenamefont {Krebs},
  \citenamefont {Taza{\"i}rt}, \citenamefont {Harouri}, \citenamefont
  {Lemaitre}, \citenamefont {Sagnes}, \citenamefont {Anton}, \citenamefont
  {Lanco}, \citenamefont {Somaschi}, \citenamefont {Loredo},\ and\
  \citenamefont {Senellart}}]{Thomas.2021b}%
  \BibitemOpen
  \bibfield  {author} {\bibinfo {author} {\bibfnamefont {S.~E.}\ \bibnamefont
  {Thomas}}, \bibinfo {author} {\bibfnamefont {M.}~\bibnamefont {Billard}},
  \bibinfo {author} {\bibfnamefont {N.}~\bibnamefont {Coste}}, \bibinfo
  {author} {\bibfnamefont {S.~C.}\ \bibnamefont {Wein}}, \bibinfo {author}
  {\bibnamefont {Priya}}, \bibinfo {author} {\bibfnamefont {H.}~\bibnamefont
  {Ollivier}}, \bibinfo {author} {\bibfnamefont {O.}~\bibnamefont {Krebs}},
  \bibinfo {author} {\bibfnamefont {L.}~\bibnamefont {Taza{\"i}rt}}, \bibinfo
  {author} {\bibfnamefont {A.}~\bibnamefont {Harouri}}, \bibinfo {author}
  {\bibfnamefont {A.}~\bibnamefont {Lemaitre}}, \bibinfo {author}
  {\bibfnamefont {I.}~\bibnamefont {Sagnes}}, \bibinfo {author} {\bibfnamefont
  {C.}~\bibnamefont {Anton}}, \bibinfo {author} {\bibfnamefont
  {L.}~\bibnamefont {Lanco}}, \bibinfo {author} {\bibfnamefont
  {N.}~\bibnamefont {Somaschi}}, \bibinfo {author} {\bibfnamefont {J.~C.}\
  \bibnamefont {Loredo}},\ and\ \bibinfo {author} {\bibfnamefont
  {P.}~\bibnamefont {Senellart}},\ }\bibfield  {title} {\bibinfo {title}
  {{Bright Polarized Single-Photon Source Based on a Linear Dipole}},\ }\href
  {https://doi.org/10.1103/PhysRevLett.126.233601} {\bibfield  {journal}
  {\bibinfo  {journal} {{Physical review letters}}\ }\textbf {\bibinfo {volume}
  {126}},\ \bibinfo {pages} {233601} (\bibinfo {year} {2021})}\BibitemShut
  {NoStop}%
\bibitem [{\citenamefont {Tomm}\ \emph {et~al.}(2021)\citenamefont {Tomm},
  \citenamefont {Javadi}, \citenamefont {Antoniadis}, \citenamefont {Najer},
  \citenamefont {L{\"o}bl}, \citenamefont {Korsch}, \citenamefont {Schott},
  \citenamefont {Valentin}, \citenamefont {Wieck}, \citenamefont {Ludwig},\
  and\ \citenamefont {Warburton}}]{Tomm.2021}%
  \BibitemOpen
  \bibfield  {author} {\bibinfo {author} {\bibfnamefont {N.}~\bibnamefont
  {Tomm}}, \bibinfo {author} {\bibfnamefont {A.}~\bibnamefont {Javadi}},
  \bibinfo {author} {\bibfnamefont {N.~O.}\ \bibnamefont {Antoniadis}},
  \bibinfo {author} {\bibfnamefont {D.}~\bibnamefont {Najer}}, \bibinfo
  {author} {\bibfnamefont {M.~C.}\ \bibnamefont {L{\"o}bl}}, \bibinfo {author}
  {\bibfnamefont {A.~R.}\ \bibnamefont {Korsch}}, \bibinfo {author}
  {\bibfnamefont {R.}~\bibnamefont {Schott}}, \bibinfo {author} {\bibfnamefont
  {S.~R.}\ \bibnamefont {Valentin}}, \bibinfo {author} {\bibfnamefont {A.~D.}\
  \bibnamefont {Wieck}}, \bibinfo {author} {\bibfnamefont {A.}~\bibnamefont
  {Ludwig}},\ and\ \bibinfo {author} {\bibfnamefont {R.~J.}\ \bibnamefont
  {Warburton}},\ }\bibfield  {title} {\bibinfo {title} {{A bright and fast
  source of coherent single photons}},\ }\href
  {https://doi.org/10.1038/s41565-020-00831-x} {\bibfield  {journal} {\bibinfo
  {journal} {{Nature Nanotechnology}}\ }\textbf {\bibinfo {volume} {16}},\
  \bibinfo {pages} {399} (\bibinfo {year} {2021})}\BibitemShut {NoStop}%
\bibitem [{\citenamefont {Gschrey}\ \emph {et~al.}(2015)\citenamefont
  {Gschrey}, \citenamefont {Thoma}, \citenamefont {Schnauber}, \citenamefont
  {Seifried}, \citenamefont {Schmidt}, \citenamefont {Wohlfeil}, \citenamefont
  {Kr{\"u}ger}, \citenamefont {Schulze}, \citenamefont {Heindel}, \citenamefont
  {Burger}, \citenamefont {Schmidt}, \citenamefont {Strittmatter},
  \citenamefont {Rodt},\ and\ \citenamefont {Reitzenstein}}]{Gschrey.2015}%
  \BibitemOpen
  \bibfield  {author} {\bibinfo {author} {\bibfnamefont {M.}~\bibnamefont
  {Gschrey}}, \bibinfo {author} {\bibfnamefont {A.}~\bibnamefont {Thoma}},
  \bibinfo {author} {\bibfnamefont {P.}~\bibnamefont {Schnauber}}, \bibinfo
  {author} {\bibfnamefont {M.}~\bibnamefont {Seifried}}, \bibinfo {author}
  {\bibfnamefont {R.}~\bibnamefont {Schmidt}}, \bibinfo {author} {\bibfnamefont
  {B.}~\bibnamefont {Wohlfeil}}, \bibinfo {author} {\bibfnamefont
  {L.}~\bibnamefont {Kr{\"u}ger}}, \bibinfo {author} {\bibfnamefont {J.-H.}\
  \bibnamefont {Schulze}}, \bibinfo {author} {\bibfnamefont {T.}~\bibnamefont
  {Heindel}}, \bibinfo {author} {\bibfnamefont {S.}~\bibnamefont {Burger}},
  \bibinfo {author} {\bibfnamefont {F.}~\bibnamefont {Schmidt}}, \bibinfo
  {author} {\bibfnamefont {A.}~\bibnamefont {Strittmatter}}, \bibinfo {author}
  {\bibfnamefont {S.}~\bibnamefont {Rodt}},\ and\ \bibinfo {author}
  {\bibfnamefont {S.}~\bibnamefont {Reitzenstein}},\ }\bibfield  {title}
  {\bibinfo {title} {{Highly indistinguishable photons from deterministic
  quantum-dot microlenses utilizing three-dimensional in situ electron-beam
  lithography}},\ }\href {https://doi.org/10.1038/ncomms8662} {\bibfield
  {journal} {\bibinfo  {journal} {{Nature communications}}\ }\textbf {\bibinfo
  {volume} {6}},\ \bibinfo {pages} {7662} (\bibinfo {year} {2015})}\BibitemShut
  {NoStop}%
\bibitem [{\citenamefont {Sartison}\ \emph {et~al.}(2017)\citenamefont
  {Sartison}, \citenamefont {Portalupi}, \citenamefont {Gissibl}, \citenamefont
  {Jetter}, \citenamefont {Giessen},\ and\ \citenamefont
  {Michler}}]{Sartison.2017}%
  \BibitemOpen
  \bibfield  {author} {\bibinfo {author} {\bibfnamefont {M.}~\bibnamefont
  {Sartison}}, \bibinfo {author} {\bibfnamefont {S.~L.}\ \bibnamefont
  {Portalupi}}, \bibinfo {author} {\bibfnamefont {T.}~\bibnamefont {Gissibl}},
  \bibinfo {author} {\bibfnamefont {M.}~\bibnamefont {Jetter}}, \bibinfo
  {author} {\bibfnamefont {H.}~\bibnamefont {Giessen}},\ and\ \bibinfo {author}
  {\bibfnamefont {P.}~\bibnamefont {Michler}},\ }\bibfield  {title} {\bibinfo
  {title} {{Combining in-situ lithography with 3D printed solid immersion
  lenses for single quantum dot spectroscopy}},\ }\href
  {https://doi.org/10.1038/srep39916} {\bibfield  {journal} {\bibinfo
  {journal} {{Scientific Reports}}\ }\textbf {\bibinfo {volume} {7}},\ \bibinfo
  {pages} {39916} (\bibinfo {year} {2017})}\BibitemShut {NoStop}%
\bibitem [{\citenamefont {Chen}\ \emph {et~al.}(2018)\citenamefont {Chen},
  \citenamefont {Zopf}, \citenamefont {Keil}, \citenamefont {Ding},\ and\
  \citenamefont {Schmidt}}]{Chen.2018}%
  \BibitemOpen
  \bibfield  {author} {\bibinfo {author} {\bibfnamefont {Y.}~\bibnamefont
  {Chen}}, \bibinfo {author} {\bibfnamefont {M.}~\bibnamefont {Zopf}}, \bibinfo
  {author} {\bibfnamefont {R.}~\bibnamefont {Keil}}, \bibinfo {author}
  {\bibfnamefont {F.}~\bibnamefont {Ding}},\ and\ \bibinfo {author}
  {\bibfnamefont {O.~G.}\ \bibnamefont {Schmidt}},\ }\bibfield  {title}
  {\bibinfo {title} {{Highly-efficient extraction of entangled photons from
  quantum dots using a broadband optical antenna}},\ }\href
  {https://doi.org/10.1038/s41467-018-05456-2} {\bibfield  {journal} {\bibinfo
  {journal} {{Nature communications}}\ }\textbf {\bibinfo {volume} {9}},\
  \bibinfo {pages} {2994} (\bibinfo {year} {2018})}\BibitemShut {NoStop}%
\bibitem [{\citenamefont {Yao}\ \emph {et~al.}(2018)\citenamefont {Yao},
  \citenamefont {Su}, \citenamefont {Wei}, \citenamefont {Liu}, \citenamefont
  {Zhao},\ and\ \citenamefont {Liu}}]{Yao.2018}%
  \BibitemOpen
  \bibfield  {author} {\bibinfo {author} {\bibfnamefont {B.}~\bibnamefont
  {Yao}}, \bibinfo {author} {\bibfnamefont {R.}~\bibnamefont {Su}}, \bibinfo
  {author} {\bibfnamefont {Y.}~\bibnamefont {Wei}}, \bibinfo {author}
  {\bibfnamefont {Z.}~\bibnamefont {Liu}}, \bibinfo {author} {\bibfnamefont
  {T.}~\bibnamefont {Zhao}},\ and\ \bibinfo {author} {\bibfnamefont
  {J.}~\bibnamefont {Liu}},\ }\bibfield  {title} {\bibinfo {title} {{Design for
  Hybrid Circular Bragg Gratings for a Highly Efficient Quantum-Dot
  Single-Photon Source}},\ }\href {https://doi.org/10.3938/jkps.73.1502}
  {\bibfield  {journal} {\bibinfo  {journal} {{Journal of the Korean Physical
  Society}}\ }\textbf {\bibinfo {volume} {73}},\ \bibinfo {pages} {1502}
  (\bibinfo {year} {2018})}\BibitemShut {NoStop}%
\bibitem [{\citenamefont {Davan{\c{c}}o}\ \emph {et~al.}(2011)\citenamefont
  {Davan{\c{c}}o}, \citenamefont {Rakher}, \citenamefont {Schuh}, \citenamefont
  {Badolato},\ and\ \citenamefont {Srinivasan}}]{Davanco.2011}%
  \BibitemOpen
  \bibfield  {author} {\bibinfo {author} {\bibfnamefont {M.}~\bibnamefont
  {Davan{\c{c}}o}}, \bibinfo {author} {\bibfnamefont {M.~T.}\ \bibnamefont
  {Rakher}}, \bibinfo {author} {\bibfnamefont {D.}~\bibnamefont {Schuh}},
  \bibinfo {author} {\bibfnamefont {A.}~\bibnamefont {Badolato}},\ and\
  \bibinfo {author} {\bibfnamefont {K.}~\bibnamefont {Srinivasan}},\ }\bibfield
   {title} {\bibinfo {title} {{A circular dielectric grating for vertical
  extraction of single quantum dot emission}},\ }\href
  {https://doi.org/10.1063/1.3615051} {\bibfield  {journal} {\bibinfo
  {journal} {{Applied Physics Letters}}\ }\textbf {\bibinfo {volume} {99}},\
  \bibinfo {pages} {041102} (\bibinfo {year} {2011})}\BibitemShut {NoStop}%
\bibitem [{\citenamefont {Sapienza}\ \emph {et~al.}(2015)\citenamefont
  {Sapienza}, \citenamefont {Davan{\c{c}}o}, \citenamefont {Badolato},\ and\
  \citenamefont {Srinivasan}}]{Sapienza.2015}%
  \BibitemOpen
  \bibfield  {author} {\bibinfo {author} {\bibfnamefont {L.}~\bibnamefont
  {Sapienza}}, \bibinfo {author} {\bibfnamefont {M.}~\bibnamefont
  {Davan{\c{c}}o}}, \bibinfo {author} {\bibfnamefont {A.}~\bibnamefont
  {Badolato}},\ and\ \bibinfo {author} {\bibfnamefont {K.}~\bibnamefont
  {Srinivasan}},\ }\bibfield  {title} {\bibinfo {title} {{Nanoscale optical
  positioning of single quantum dots for bright and pure single-photon
  emission}},\ }\href {https://doi.org/10.1038/ncomms8833} {\bibfield
  {journal} {\bibinfo  {journal} {{Nature communications}}\ }\textbf {\bibinfo
  {volume} {6}},\ \bibinfo {pages} {7833 EP } (\bibinfo {year}
  {2015})}\BibitemShut {NoStop}%
\bibitem [{\citenamefont {Kolatschek}\ \emph {et~al.}(2019)\citenamefont
  {Kolatschek}, \citenamefont {Hepp}, \citenamefont {Sartison}, \citenamefont
  {Jetter}, \citenamefont {Michler},\ and\ \citenamefont
  {Portalupi}}]{Kolatschek.2019}%
  \BibitemOpen
  \bibfield  {author} {\bibinfo {author} {\bibfnamefont {S.}~\bibnamefont
  {Kolatschek}}, \bibinfo {author} {\bibfnamefont {S.}~\bibnamefont {Hepp}},
  \bibinfo {author} {\bibfnamefont {M.}~\bibnamefont {Sartison}}, \bibinfo
  {author} {\bibfnamefont {M.}~\bibnamefont {Jetter}}, \bibinfo {author}
  {\bibfnamefont {P.}~\bibnamefont {Michler}},\ and\ \bibinfo {author}
  {\bibfnamefont {S.~L.}\ \bibnamefont {Portalupi}},\ }\bibfield  {title}
  {\bibinfo {title} {{Deterministic fabrication of circular Bragg gratings
  coupled to single quantum emitters via the combination of in-situ optical
  lithography and electron-beam lithography}},\ }\href
  {https://doi.org/10.1063/1.5050344} {\bibfield  {journal} {\bibinfo
  {journal} {{Journal of Applied Physics}}\ }\textbf {\bibinfo {volume}
  {125}},\ \bibinfo {pages} {045701} (\bibinfo {year} {2019})}\BibitemShut
  {NoStop}%
\bibitem [{\citenamefont {Liu}\ \emph {et~al.}(2019)\citenamefont {Liu},
  \citenamefont {Su}, \citenamefont {Wei}, \citenamefont {Yao}, \citenamefont
  {Silva}, \citenamefont {Yu}, \citenamefont {Iles-Smith}, \citenamefont
  {Srinivasan}, \citenamefont {Rastelli}, \citenamefont {Li},\ and\
  \citenamefont {Wang}}]{Liu.2019}%
  \BibitemOpen
  \bibfield  {author} {\bibinfo {author} {\bibfnamefont {J.}~\bibnamefont
  {Liu}}, \bibinfo {author} {\bibfnamefont {R.}~\bibnamefont {Su}}, \bibinfo
  {author} {\bibfnamefont {Y.}~\bibnamefont {Wei}}, \bibinfo {author}
  {\bibfnamefont {B.}~\bibnamefont {Yao}}, \bibinfo {author} {\bibfnamefont
  {S.~F. C.~d.}\ \bibnamefont {Silva}}, \bibinfo {author} {\bibfnamefont
  {Y.}~\bibnamefont {Yu}}, \bibinfo {author} {\bibfnamefont {J.}~\bibnamefont
  {Iles-Smith}}, \bibinfo {author} {\bibfnamefont {K.}~\bibnamefont
  {Srinivasan}}, \bibinfo {author} {\bibfnamefont {A.}~\bibnamefont
  {Rastelli}}, \bibinfo {author} {\bibfnamefont {J.}~\bibnamefont {Li}},\ and\
  \bibinfo {author} {\bibfnamefont {X.}~\bibnamefont {Wang}},\ }\bibfield
  {title} {\bibinfo {title} {{A solid-state source of strongly entangled photon
  pairs with high brightness and indistinguishability}},\ }\href
  {https://doi.org/10.1038/s41565-019-0435-9} {\bibfield  {journal} {\bibinfo
  {journal} {{Nature Nanotechnology}}\ }\textbf {\bibinfo {volume} {14}},\
  \bibinfo {pages} {586} (\bibinfo {year} {2019})}\BibitemShut {NoStop}%
\bibitem [{\citenamefont {Wang}\ \emph
  {et~al.}(2019{\natexlab{a}})\citenamefont {Wang}, \citenamefont {Hu},
  \citenamefont {Chung}, \citenamefont {Qin}, \citenamefont {Yang},
  \citenamefont {Li}, \citenamefont {Liu}, \citenamefont {Zhong}, \citenamefont
  {He}, \citenamefont {Ding}, \citenamefont {Deng}, \citenamefont {Dai},
  \citenamefont {Huo}, \citenamefont {H{\"o}fling}, \citenamefont {Lu},\ and\
  \citenamefont {Pan}}]{Wang.2019}%
  \BibitemOpen
  \bibfield  {author} {\bibinfo {author} {\bibfnamefont {H.}~\bibnamefont
  {Wang}}, \bibinfo {author} {\bibfnamefont {H.}~\bibnamefont {Hu}}, \bibinfo
  {author} {\bibfnamefont {T.-H.}\ \bibnamefont {Chung}}, \bibinfo {author}
  {\bibfnamefont {J.}~\bibnamefont {Qin}}, \bibinfo {author} {\bibfnamefont
  {X.}~\bibnamefont {Yang}}, \bibinfo {author} {\bibfnamefont {J.-P.}\
  \bibnamefont {Li}}, \bibinfo {author} {\bibfnamefont {R.-Z.}\ \bibnamefont
  {Liu}}, \bibinfo {author} {\bibfnamefont {H.-S.}\ \bibnamefont {Zhong}},
  \bibinfo {author} {\bibfnamefont {Y.-M.}\ \bibnamefont {He}}, \bibinfo
  {author} {\bibfnamefont {X.}~\bibnamefont {Ding}}, \bibinfo {author}
  {\bibfnamefont {Y.-H.}\ \bibnamefont {Deng}}, \bibinfo {author}
  {\bibfnamefont {Q.}~\bibnamefont {Dai}}, \bibinfo {author} {\bibfnamefont
  {Y.-H.}\ \bibnamefont {Huo}}, \bibinfo {author} {\bibfnamefont
  {S.}~\bibnamefont {H{\"o}fling}}, \bibinfo {author} {\bibfnamefont {C.-Y.}\
  \bibnamefont {Lu}},\ and\ \bibinfo {author} {\bibfnamefont {J.-W.}\
  \bibnamefont {Pan}},\ }\bibfield  {title} {\bibinfo {title} {{On-Demand
  Semiconductor Source of Entangled Photons Which Simultaneously Has High
  Fidelity, Efficiency, and Indistinguishability}},\ }\href
  {https://doi.org/10.1103/PhysRevLett.122.113602} {\bibfield  {journal}
  {\bibinfo  {journal} {{Physical review letters}}\ }\textbf {\bibinfo {volume}
  {122}},\ \bibinfo {pages} {113602} (\bibinfo {year}
  {2019}{\natexlab{a}})}\BibitemShut {NoStop}%
\bibitem [{\citenamefont {Wang}\ \emph
  {et~al.}(2019{\natexlab{b}})\citenamefont {Wang}, \citenamefont {He},
  \citenamefont {Chung}, \citenamefont {Hu}, \citenamefont {Yu}, \citenamefont
  {Chen}, \citenamefont {Ding}, \citenamefont {Chen}, \citenamefont {Qin},
  \citenamefont {Yang}, \citenamefont {Liu}, \citenamefont {Duan},
  \citenamefont {Li}, \citenamefont {Gerhardt}, \citenamefont {Winkler},
  \citenamefont {Jurkat}, \citenamefont {Wang}, \citenamefont {Gregersen},
  \citenamefont {Huo}, \citenamefont {Dai}, \citenamefont {Yu}, \citenamefont
  {H{\"o}fling}, \citenamefont {Lu},\ and\ \citenamefont {Pan}}]{Wang.2019c}%
  \BibitemOpen
  \bibfield  {author} {\bibinfo {author} {\bibfnamefont {H.}~\bibnamefont
  {Wang}}, \bibinfo {author} {\bibfnamefont {Y.-M.}\ \bibnamefont {He}},
  \bibinfo {author} {\bibfnamefont {T.-H.}\ \bibnamefont {Chung}}, \bibinfo
  {author} {\bibfnamefont {H.}~\bibnamefont {Hu}}, \bibinfo {author}
  {\bibfnamefont {Y.}~\bibnamefont {Yu}}, \bibinfo {author} {\bibfnamefont
  {S.}~\bibnamefont {Chen}}, \bibinfo {author} {\bibfnamefont {X.}~\bibnamefont
  {Ding}}, \bibinfo {author} {\bibfnamefont {M.-C.}\ \bibnamefont {Chen}},
  \bibinfo {author} {\bibfnamefont {J.}~\bibnamefont {Qin}}, \bibinfo {author}
  {\bibfnamefont {X.}~\bibnamefont {Yang}}, \bibinfo {author} {\bibfnamefont
  {R.-Z.}\ \bibnamefont {Liu}}, \bibinfo {author} {\bibfnamefont {Z.-C.}\
  \bibnamefont {Duan}}, \bibinfo {author} {\bibfnamefont {J.-P.}\ \bibnamefont
  {Li}}, \bibinfo {author} {\bibfnamefont {S.}~\bibnamefont {Gerhardt}},
  \bibinfo {author} {\bibfnamefont {K.}~\bibnamefont {Winkler}}, \bibinfo
  {author} {\bibfnamefont {J.}~\bibnamefont {Jurkat}}, \bibinfo {author}
  {\bibfnamefont {L.-J.}\ \bibnamefont {Wang}}, \bibinfo {author}
  {\bibfnamefont {N.}~\bibnamefont {Gregersen}}, \bibinfo {author}
  {\bibfnamefont {Y.-H.}\ \bibnamefont {Huo}}, \bibinfo {author} {\bibfnamefont
  {Q.}~\bibnamefont {Dai}}, \bibinfo {author} {\bibfnamefont {S.}~\bibnamefont
  {Yu}}, \bibinfo {author} {\bibfnamefont {S.}~\bibnamefont {H{\"o}fling}},
  \bibinfo {author} {\bibfnamefont {C.-Y.}\ \bibnamefont {Lu}},\ and\ \bibinfo
  {author} {\bibfnamefont {J.-W.}\ \bibnamefont {Pan}},\ }\bibfield  {title}
  {\bibinfo {title} {{Towards optimal single-photon sources from polarized
  microcavities}},\ }\href {https://doi.org/10.1038/s41566-019-0494-3}
  {\bibfield  {journal} {\bibinfo  {journal} {{Nature Photonics}}\ }\textbf
  {\bibinfo {volume} {13}},\ \bibinfo {pages} {770} (\bibinfo {year}
  {2019}{\natexlab{b}})}\BibitemShut {NoStop}%
\bibitem [{\citenamefont {You}\ \emph {et~al.}()\citenamefont {You},
  \citenamefont {Zheng}, \citenamefont {Chen}, \citenamefont {Liu},
  \citenamefont {Qin}, \citenamefont {Xu}, \citenamefont {Ge}, \citenamefont
  {Chung}, \citenamefont {Qiao}, \citenamefont {Jiang}, \citenamefont {Zhong},
  \citenamefont {Chen}, \citenamefont {Wang}, \citenamefont {He}, \citenamefont
  {Xie}, \citenamefont {Li}, \citenamefont {You}, \citenamefont {Schneider},
  \citenamefont {Yin}, \citenamefont {Chen}, \citenamefont {Benyoucef},
  \citenamefont {Huo}, \citenamefont {Hoefling}, \citenamefont {Zhang},
  \citenamefont {Lu},\ and\ \citenamefont {Pan}}]{You.2021}%
  \BibitemOpen
  \bibfield  {author} {\bibinfo {author} {\bibfnamefont {X.}~\bibnamefont
  {You}}, \bibinfo {author} {\bibfnamefont {M.-Y.}\ \bibnamefont {Zheng}},
  \bibinfo {author} {\bibfnamefont {S.}~\bibnamefont {Chen}}, \bibinfo {author}
  {\bibfnamefont {R.-Z.}\ \bibnamefont {Liu}}, \bibinfo {author} {\bibfnamefont
  {J.}~\bibnamefont {Qin}}, \bibinfo {author} {\bibfnamefont {M.-C.}\
  \bibnamefont {Xu}}, \bibinfo {author} {\bibfnamefont {Z.-X.}\ \bibnamefont
  {Ge}}, \bibinfo {author} {\bibfnamefont {T.-H.}\ \bibnamefont {Chung}},
  \bibinfo {author} {\bibfnamefont {Y.-K.}\ \bibnamefont {Qiao}}, \bibinfo
  {author} {\bibfnamefont {Y.-F.}\ \bibnamefont {Jiang}}, \bibinfo {author}
  {\bibfnamefont {H.-S.}\ \bibnamefont {Zhong}}, \bibinfo {author}
  {\bibfnamefont {M.-C.}\ \bibnamefont {Chen}}, \bibinfo {author}
  {\bibfnamefont {H.}~\bibnamefont {Wang}}, \bibinfo {author} {\bibfnamefont
  {Y.-M.}\ \bibnamefont {He}}, \bibinfo {author} {\bibfnamefont {X.-P.}\
  \bibnamefont {Xie}}, \bibinfo {author} {\bibfnamefont {H.}~\bibnamefont
  {Li}}, \bibinfo {author} {\bibfnamefont {L.-X.}\ \bibnamefont {You}},
  \bibinfo {author} {\bibfnamefont {C.}~\bibnamefont {Schneider}}, \bibinfo
  {author} {\bibfnamefont {J.}~\bibnamefont {Yin}}, \bibinfo {author}
  {\bibfnamefont {T.-Y.}\ \bibnamefont {Chen}}, \bibinfo {author}
  {\bibfnamefont {M.}~\bibnamefont {Benyoucef}}, \bibinfo {author}
  {\bibfnamefont {Y.-H.}\ \bibnamefont {Huo}}, \bibinfo {author} {\bibfnamefont
  {S.}~\bibnamefont {Hoefling}}, \bibinfo {author} {\bibfnamefont
  {Q.}~\bibnamefont {Zhang}}, \bibinfo {author} {\bibfnamefont {C.-Y.}\
  \bibnamefont {Lu}},\ and\ \bibinfo {author} {\bibfnamefont {J.-W.}\
  \bibnamefont {Pan}},\ }\bibfield  {title} {\bibinfo {title} {{Quantum
  interference between independent solid-state single-photon sources separated
  by 300 km fiber}},\ }\bibfield  {journal} {\bibinfo  {journal}
  {{arXiv.2106.15545}}\ }\href {https://doi.org/10.48550/arXiv.2106.15545}
  {10.48550/arXiv.2106.15545}\BibitemShut {NoStop}%
\bibitem [{\citenamefont {Rickert}\ \emph {et~al.}(2019)\citenamefont
  {Rickert}, \citenamefont {Kupko}, \citenamefont {Rodt}, \citenamefont
  {Reitzenstein},\ and\ \citenamefont {Heindel}}]{Rickert.2019}%
  \BibitemOpen
  \bibfield  {author} {\bibinfo {author} {\bibfnamefont {L.}~\bibnamefont
  {Rickert}}, \bibinfo {author} {\bibfnamefont {T.}~\bibnamefont {Kupko}},
  \bibinfo {author} {\bibfnamefont {S.}~\bibnamefont {Rodt}}, \bibinfo {author}
  {\bibfnamefont {S.}~\bibnamefont {Reitzenstein}},\ and\ \bibinfo {author}
  {\bibfnamefont {T.}~\bibnamefont {Heindel}},\ }\bibfield  {title} {\bibinfo
  {title} {{Optimized designs for telecom-wavelength quantum light sources
  based on hybrid circular Bragg gratings}},\ }\href
  {https://doi.org/10.1364/OE.27.036824} {\bibfield  {journal} {\bibinfo
  {journal} {{Optics express}}\ }\textbf {\bibinfo {volume} {27}},\ \bibinfo
  {pages} {36824} (\bibinfo {year} {2019})}\BibitemShut {NoStop}%
\bibitem [{\citenamefont {Blokhin}\ \emph {et~al.}(2021)\citenamefont
  {Blokhin}, \citenamefont {Bobrov}, \citenamefont {Maleev}, \citenamefont
  {Donges}, \citenamefont {Bremer}, \citenamefont {Blokhin}, \citenamefont
  {Vasil'ev}, \citenamefont {Kuzmenkov}, \citenamefont {Kolodeznyi},
  \citenamefont {Shchukin}, \citenamefont {Ledentsov}, \citenamefont
  {Reitzenstein},\ and\ \citenamefont {Ustinov}}]{Blokhin.2021}%
  \BibitemOpen
  \bibfield  {author} {\bibinfo {author} {\bibfnamefont {S.~A.}\ \bibnamefont
  {Blokhin}}, \bibinfo {author} {\bibfnamefont {M.~A.}\ \bibnamefont {Bobrov}},
  \bibinfo {author} {\bibfnamefont {N.~A.}\ \bibnamefont {Maleev}}, \bibinfo
  {author} {\bibfnamefont {J.~N.}\ \bibnamefont {Donges}}, \bibinfo {author}
  {\bibfnamefont {L.}~\bibnamefont {Bremer}}, \bibinfo {author} {\bibfnamefont
  {A.~A.}\ \bibnamefont {Blokhin}}, \bibinfo {author} {\bibfnamefont {A.~P.}\
  \bibnamefont {Vasil'ev}}, \bibinfo {author} {\bibfnamefont {A.~G.}\
  \bibnamefont {Kuzmenkov}}, \bibinfo {author} {\bibfnamefont {E.~S.}\
  \bibnamefont {Kolodeznyi}}, \bibinfo {author} {\bibfnamefont {V.~A.}\
  \bibnamefont {Shchukin}}, \bibinfo {author} {\bibfnamefont {N.~N.}\
  \bibnamefont {Ledentsov}}, \bibinfo {author} {\bibfnamefont {S.}~\bibnamefont
  {Reitzenstein}},\ and\ \bibinfo {author} {\bibfnamefont {V.~M.}\ \bibnamefont
  {Ustinov}},\ }\bibfield  {title} {\bibinfo {title} {{Design optimization for
  bright electrically-driven quantum dot single-photon sources emitting in
  telecom O-band}},\ }\href {https://doi.org/10.1364/OE.415979} {\bibfield
  {journal} {\bibinfo  {journal} {{Optics express}}\ }\textbf {\bibinfo
  {volume} {29}},\ \bibinfo {pages} {6582} (\bibinfo {year}
  {2021})}\BibitemShut {NoStop}%
\bibitem [{\citenamefont {Barbiero}\ \emph {et~al.}(2022)\citenamefont
  {Barbiero}, \citenamefont {Huwer}, \citenamefont {Skiba-Szymanska},
  \citenamefont {M{\"u}ller}, \citenamefont {Stevenson},\ and\ \citenamefont
  {Shields}}]{Barbiero.2022}%
  \BibitemOpen
  \bibfield  {author} {\bibinfo {author} {\bibfnamefont {A.}~\bibnamefont
  {Barbiero}}, \bibinfo {author} {\bibfnamefont {J.}~\bibnamefont {Huwer}},
  \bibinfo {author} {\bibfnamefont {J.}~\bibnamefont {Skiba-Szymanska}},
  \bibinfo {author} {\bibfnamefont {T.}~\bibnamefont {M{\"u}ller}}, \bibinfo
  {author} {\bibfnamefont {R.~M.}\ \bibnamefont {Stevenson}},\ and\ \bibinfo
  {author} {\bibfnamefont {A.~J.}\ \bibnamefont {Shields}},\ }\bibfield
  {title} {\bibinfo {title} {{Design study for an efficient semiconductor
  quantum light source operating in the telecom C-band based on an
  electrically-driven circular Bragg grating}},\ }\href
  {https://doi.org/10.1364/OE.452328} {\bibfield  {journal} {\bibinfo
  {journal} {{Optics express}}\ }\textbf {\bibinfo {volume} {30}},\ \bibinfo
  {pages} {10919} (\bibinfo {year} {2022})}\BibitemShut {NoStop}%
\bibitem [{\citenamefont {Bremer}\ \emph {et~al.}(2022)\citenamefont {Bremer},
  \citenamefont {Jimenez}, \citenamefont {Thiele}, \citenamefont {Weber},
  \citenamefont {Huber}, \citenamefont {Rodt}, \citenamefont {Herkommer},
  \citenamefont {Burger}, \citenamefont {H{\"o}fling}, \citenamefont
  {Giessen},\ and\ \citenamefont {Reitzenstein}}]{Bremer.2022}%
  \BibitemOpen
  \bibfield  {author} {\bibinfo {author} {\bibfnamefont {L.}~\bibnamefont
  {Bremer}}, \bibinfo {author} {\bibfnamefont {C.}~\bibnamefont {Jimenez}},
  \bibinfo {author} {\bibfnamefont {S.}~\bibnamefont {Thiele}}, \bibinfo
  {author} {\bibfnamefont {K.}~\bibnamefont {Weber}}, \bibinfo {author}
  {\bibfnamefont {T.}~\bibnamefont {Huber}}, \bibinfo {author} {\bibfnamefont
  {S.}~\bibnamefont {Rodt}}, \bibinfo {author} {\bibfnamefont {A.}~\bibnamefont
  {Herkommer}}, \bibinfo {author} {\bibfnamefont {S.}~\bibnamefont {Burger}},
  \bibinfo {author} {\bibfnamefont {S.}~\bibnamefont {H{\"o}fling}}, \bibinfo
  {author} {\bibfnamefont {H.}~\bibnamefont {Giessen}},\ and\ \bibinfo {author}
  {\bibfnamefont {S.}~\bibnamefont {Reitzenstein}},\ }\bibfield  {title}
  {\bibinfo {title} {{Numerical optimization of single-mode fiber-coupled
  single-photon sources based on semiconductor quantum dots}},\ }\href
  {https://doi.org/10.1364/OE.456777} {\bibfield  {journal} {\bibinfo
  {journal} {{Optics express}}\ }\textbf {\bibinfo {volume} {30}},\ \bibinfo
  {pages} {15913} (\bibinfo {year} {2022})}\BibitemShut {NoStop}%
\bibitem [{\citenamefont {Kolatschek}\ \emph {et~al.}(2021)\citenamefont
  {Kolatschek}, \citenamefont {Nawrath}, \citenamefont {Bauer}, \citenamefont
  {Huang}, \citenamefont {Fischer}, \citenamefont {Sittig}, \citenamefont
  {Jetter}, \citenamefont {Portalupi},\ and\ \citenamefont
  {Michler}}]{Kolatschek.2021}%
  \BibitemOpen
  \bibfield  {author} {\bibinfo {author} {\bibfnamefont {S.}~\bibnamefont
  {Kolatschek}}, \bibinfo {author} {\bibfnamefont {C.}~\bibnamefont {Nawrath}},
  \bibinfo {author} {\bibfnamefont {S.}~\bibnamefont {Bauer}}, \bibinfo
  {author} {\bibfnamefont {J.}~\bibnamefont {Huang}}, \bibinfo {author}
  {\bibfnamefont {J.}~\bibnamefont {Fischer}}, \bibinfo {author} {\bibfnamefont
  {R.}~\bibnamefont {Sittig}}, \bibinfo {author} {\bibfnamefont
  {M.}~\bibnamefont {Jetter}}, \bibinfo {author} {\bibfnamefont {S.~L.}\
  \bibnamefont {Portalupi}},\ and\ \bibinfo {author} {\bibfnamefont
  {P.}~\bibnamefont {Michler}},\ }\bibfield  {title} {\bibinfo {title} {{Bright
  Purcell Enhanced Single-Photon Source in the Telecom O-Band Based on a
  Quantum Dot in a Circular Bragg Grating}},\ }\href
  {https://doi.org/10.1021/acs.nanolett.1c02647} {\bibfield  {journal}
  {\bibinfo  {journal} {{Nano letters}}\ }\textbf {\bibinfo {volume} {21}},\
  \bibinfo {pages} {7740} (\bibinfo {year} {2021})}\BibitemShut {NoStop}%
\bibitem [{\citenamefont {Reinert}(2015)}]{Reinert.2015}%
  \BibitemOpen
  \bibfield  {author} {\bibinfo {author} {\bibfnamefont {W.}~\bibnamefont
  {Reinert}},\ }\href {https://doi.org/10.1016/C2013-0-19270-7} {\emph
  {\bibinfo {title} {{Handbook of Silicon Based MEMS Materials and
  Technologies}}}}\ (\bibinfo  {publisher} {Elsevier},\ \bibinfo {year}
  {2015})\BibitemShut {NoStop}%
\bibitem [{\citenamefont {DeSalvo}\ \emph {et~al.}(1992)\citenamefont
  {DeSalvo}, \citenamefont {Tseng},\ and\ \citenamefont
  {Comas}}]{DeSalvo.1992}%
  \BibitemOpen
  \bibfield  {author} {\bibinfo {author} {\bibfnamefont {G.~C.}\ \bibnamefont
  {DeSalvo}}, \bibinfo {author} {\bibfnamefont {W.~F.}\ \bibnamefont {Tseng}},\
  and\ \bibinfo {author} {\bibfnamefont {J.}~\bibnamefont {Comas}},\ }\bibfield
   {title} {\bibinfo {title} {{Etch Rates and Selectivities of Citric
  Acid/Hydrogen Peroxide on GaAs , Al\textsubscript{0.3}Ga\textsubscript{0.7}As
  , In\textsubscript{0.2}Ga\textsubscript{0.8}As ,
  In\textsubscript{0.53}Ga\textsubscript{0.47}As ,
  In\textsubscript{0.52}Al\textsubscript{0.48}As , and InP}},\ }\href
  {https://doi.org/10.1149/1.2069311} {\bibfield  {journal} {\bibinfo
  {journal} {{Journal of The Electrochemical Society}}\ }\textbf {\bibinfo
  {volume} {139}},\ \bibinfo {pages} {831} (\bibinfo {year}
  {1992})}\BibitemShut {NoStop}%
\bibitem [{\citenamefont {Moon}\ \emph {et~al.}(1998)\citenamefont {Moon},
  \citenamefont {Lee},\ and\ \citenamefont {Yoo}}]{Moon.1998}%
  \BibitemOpen
  \bibfield  {author} {\bibinfo {author} {\bibfnamefont {E.-A.}\ \bibnamefont
  {Moon}}, \bibinfo {author} {\bibfnamefont {J.-L.}\ \bibnamefont {Lee}},\ and\
  \bibinfo {author} {\bibfnamefont {H.~M.}\ \bibnamefont {Yoo}},\ }\bibfield
  {title} {\bibinfo {title} {{Selective wet etching of GaAs on
  Al\textsubscript{x}Ga\textsubscript{1$-$x}As for AlGaAs/InGaAs/AlGaAs
  pseudomorphic high electron mobility transistor}},\ }\href
  {https://doi.org/10.1063/1.368571} {\bibfield  {journal} {\bibinfo  {journal}
  {{Journal of Applied Physics}}\ }\textbf {\bibinfo {volume} {84}},\ \bibinfo
  {pages} {3933} (\bibinfo {year} {1998})}\BibitemShut {NoStop}%
\bibitem [{\citenamefont {Fischbach}\ \emph {et~al.}(2017)\citenamefont
  {Fischbach}, \citenamefont {Kaganskiy}, \citenamefont {Tauscher},
  \citenamefont {Gericke}, \citenamefont {Thoma}, \citenamefont {Schmidt},
  \citenamefont {Strittmatter}, \citenamefont {Heindel}, \citenamefont {Rodt},\
  and\ \citenamefont {Reitzenstein}}]{Fischbach.2017}%
  \BibitemOpen
  \bibfield  {author} {\bibinfo {author} {\bibfnamefont {S.}~\bibnamefont
  {Fischbach}}, \bibinfo {author} {\bibfnamefont {A.}~\bibnamefont
  {Kaganskiy}}, \bibinfo {author} {\bibfnamefont {E.~B.~Y.}\ \bibnamefont
  {Tauscher}}, \bibinfo {author} {\bibfnamefont {F.}~\bibnamefont {Gericke}},
  \bibinfo {author} {\bibfnamefont {A.}~\bibnamefont {Thoma}}, \bibinfo
  {author} {\bibfnamefont {R.}~\bibnamefont {Schmidt}}, \bibinfo {author}
  {\bibfnamefont {A.}~\bibnamefont {Strittmatter}}, \bibinfo {author}
  {\bibfnamefont {T.}~\bibnamefont {Heindel}}, \bibinfo {author} {\bibfnamefont
  {S.}~\bibnamefont {Rodt}},\ and\ \bibinfo {author} {\bibfnamefont
  {S.}~\bibnamefont {Reitzenstein}},\ }\bibfield  {title} {\bibinfo {title}
  {{Efficient single-photon source based on a deterministically fabricated
  single quantum dot - microstructure with backside gold mirror}},\ }\href
  {https://doi.org/10.1063/1.4991389} {\bibfield  {journal} {\bibinfo
  {journal} {{Applied Physics Letters}}\ }\textbf {\bibinfo {volume} {111}},\
  \bibinfo {pages} {011106} (\bibinfo {year} {2017})}\BibitemShut {NoStop}%
\bibitem [{\citenamefont {Schmidt}\ \emph {et~al.}(2020)\citenamefont
  {Schmidt}, \citenamefont {Helversen}, \citenamefont {Fischbach},
  \citenamefont {Kaganskiy}, \citenamefont {Schmidt}, \citenamefont {Schliwa},
  \citenamefont {Heindel}, \citenamefont {Rodt},\ and\ \citenamefont
  {Reitzenstein}}]{Schmidt.2020}%
  \BibitemOpen
  \bibfield  {author} {\bibinfo {author} {\bibfnamefont {M.}~\bibnamefont
  {Schmidt}}, \bibinfo {author} {\bibfnamefont {M.~V.}\ \bibnamefont
  {Helversen}}, \bibinfo {author} {\bibfnamefont {S.}~\bibnamefont
  {Fischbach}}, \bibinfo {author} {\bibfnamefont {A.}~\bibnamefont
  {Kaganskiy}}, \bibinfo {author} {\bibfnamefont {R.}~\bibnamefont {Schmidt}},
  \bibinfo {author} {\bibfnamefont {A.}~\bibnamefont {Schliwa}}, \bibinfo
  {author} {\bibfnamefont {T.}~\bibnamefont {Heindel}}, \bibinfo {author}
  {\bibfnamefont {S.}~\bibnamefont {Rodt}},\ and\ \bibinfo {author}
  {\bibfnamefont {S.}~\bibnamefont {Reitzenstein}},\ }\bibfield  {title}
  {\bibinfo {title} {{Deterministically fabricated spectrally-tunable quantum
  dot based single-photon source}},\ }\href
  {https://doi.org/10.1364/OME.10.000076} {\bibfield  {journal} {\bibinfo
  {journal} {{Optical Materials Express}}\ }\textbf {\bibinfo {volume} {10}},\
  \bibinfo {pages} {76} (\bibinfo {year} {2020})}\BibitemShut {NoStop}%
\bibitem [{\citenamefont {Srocka}\ \emph {et~al.}(2020)\citenamefont {Srocka},
  \citenamefont {Mrowi{\'n}ski}, \citenamefont {Gro{\ss}e}, \citenamefont {von
  Helversen}, \citenamefont {Heindel}, \citenamefont {Rodt},\ and\
  \citenamefont {Reitzenstein}}]{Srocka.2020}%
  \BibitemOpen
  \bibfield  {author} {\bibinfo {author} {\bibfnamefont {N.}~\bibnamefont
  {Srocka}}, \bibinfo {author} {\bibfnamefont {P.}~\bibnamefont
  {Mrowi{\'n}ski}}, \bibinfo {author} {\bibfnamefont {J.}~\bibnamefont
  {Gro{\ss}e}}, \bibinfo {author} {\bibfnamefont {M.}~\bibnamefont {von
  Helversen}}, \bibinfo {author} {\bibfnamefont {T.}~\bibnamefont {Heindel}},
  \bibinfo {author} {\bibfnamefont {S.}~\bibnamefont {Rodt}},\ and\ \bibinfo
  {author} {\bibfnamefont {S.}~\bibnamefont {Reitzenstein}},\ }\bibfield
  {title} {\bibinfo {title} {{Deterministically fabricated quantum dot
  single-photon source emitting indistinguishable photons in the telecom
  O-band}},\ }\href {https://doi.org/10.1063/5.0010436} {\bibfield  {journal}
  {\bibinfo  {journal} {{Applied Physics Letters}}\ }\textbf {\bibinfo {volume}
  {116}},\ \bibinfo {pages} {231104} (\bibinfo {year} {2020})}\BibitemShut
  {NoStop}%
\bibitem [{\citenamefont {Kim}\ \emph {et~al.}(2016)\citenamefont {Kim},
  \citenamefont {Cai}, \citenamefont {Richardson}, \citenamefont {Leavitt},\
  and\ \citenamefont {Waks}}]{Kim.2016}%
  \BibitemOpen
  \bibfield  {author} {\bibinfo {author} {\bibfnamefont {J.-H.}\ \bibnamefont
  {Kim}}, \bibinfo {author} {\bibfnamefont {T.}~\bibnamefont {Cai}}, \bibinfo
  {author} {\bibfnamefont {C.~J.~K.}\ \bibnamefont {Richardson}}, \bibinfo
  {author} {\bibfnamefont {R.~P.}\ \bibnamefont {Leavitt}},\ and\ \bibinfo
  {author} {\bibfnamefont {E.}~\bibnamefont {Waks}},\ }\bibfield  {title}
  {\bibinfo {title} {{Two-photon interference from a bright single-photon
  source at telecom wavelengths}},\ }\href
  {https://doi.org/10.1364/OPTICA.3.000577} {\bibfield  {journal} {\bibinfo
  {journal} {{Optica}}\ }\textbf {\bibinfo {volume} {3}},\ \bibinfo {pages}
  {577} (\bibinfo {year} {2016})}\BibitemShut {NoStop}%
\bibitem [{\citenamefont {Yang}\ \emph {et~al.}(2020)\citenamefont {Yang},
  \citenamefont {Nawrath}, \citenamefont {Keil}, \citenamefont {Joos},
  \citenamefont {Zhang}, \citenamefont {H{\"o}fer}, \citenamefont {Chen},
  \citenamefont {Zopf}, \citenamefont {Jetter}, \citenamefont {{Luca
  Portalupi}}, \citenamefont {Ding}, \citenamefont {Michler},\ and\
  \citenamefont {Schmidt}}]{Yang.2020}%
  \BibitemOpen
  \bibfield  {author} {\bibinfo {author} {\bibfnamefont {J.}~\bibnamefont
  {Yang}}, \bibinfo {author} {\bibfnamefont {C.}~\bibnamefont {Nawrath}},
  \bibinfo {author} {\bibfnamefont {R.}~\bibnamefont {Keil}}, \bibinfo {author}
  {\bibfnamefont {R.}~\bibnamefont {Joos}}, \bibinfo {author} {\bibfnamefont
  {X.}~\bibnamefont {Zhang}}, \bibinfo {author} {\bibfnamefont
  {B.}~\bibnamefont {H{\"o}fer}}, \bibinfo {author} {\bibfnamefont
  {Y.}~\bibnamefont {Chen}}, \bibinfo {author} {\bibfnamefont {M.}~\bibnamefont
  {Zopf}}, \bibinfo {author} {\bibfnamefont {M.}~\bibnamefont {Jetter}},
  \bibinfo {author} {\bibfnamefont {S.}~\bibnamefont {{Luca Portalupi}}},
  \bibinfo {author} {\bibfnamefont {F.}~\bibnamefont {Ding}}, \bibinfo {author}
  {\bibfnamefont {P.}~\bibnamefont {Michler}},\ and\ \bibinfo {author}
  {\bibfnamefont {O.~G.}\ \bibnamefont {Schmidt}},\ }\bibfield  {title}
  {\bibinfo {title} {{Quantum dot-based broadband optical antenna for efficient
  extraction of single photons in the telecom O-band}},\ }\href
  {https://doi.org/10.1364/OE.395367} {\bibfield  {journal} {\bibinfo
  {journal} {{Optics express}}\ }\textbf {\bibinfo {volume} {28}},\ \bibinfo
  {pages} {19457} (\bibinfo {year} {2020})}\BibitemShut {NoStop}%
\bibitem [{\citenamefont {Morrison}\ \emph {et~al.}(2021)\citenamefont
  {Morrison}, \citenamefont {Rambach}, \citenamefont {Koong}, \citenamefont
  {Graffitti}, \citenamefont {Thorburn}, \citenamefont {Kar}, \citenamefont
  {Ma}, \citenamefont {Park}, \citenamefont {Song}, \citenamefont {Stoltz},
  \citenamefont {Bouwmeester}, \citenamefont {Fedrizzi},\ and\ \citenamefont
  {Gerardot}}]{Morrison.2021}%
  \BibitemOpen
  \bibfield  {author} {\bibinfo {author} {\bibfnamefont {C.~L.}\ \bibnamefont
  {Morrison}}, \bibinfo {author} {\bibfnamefont {M.}~\bibnamefont {Rambach}},
  \bibinfo {author} {\bibfnamefont {Z.~X.}\ \bibnamefont {Koong}}, \bibinfo
  {author} {\bibfnamefont {F.}~\bibnamefont {Graffitti}}, \bibinfo {author}
  {\bibfnamefont {F.}~\bibnamefont {Thorburn}}, \bibinfo {author}
  {\bibfnamefont {A.~K.}\ \bibnamefont {Kar}}, \bibinfo {author} {\bibfnamefont
  {Y.}~\bibnamefont {Ma}}, \bibinfo {author} {\bibfnamefont {S.-I.}\
  \bibnamefont {Park}}, \bibinfo {author} {\bibfnamefont {J.~D.}\ \bibnamefont
  {Song}}, \bibinfo {author} {\bibfnamefont {N.~G.}\ \bibnamefont {Stoltz}},
  \bibinfo {author} {\bibfnamefont {D.}~\bibnamefont {Bouwmeester}}, \bibinfo
  {author} {\bibfnamefont {A.}~\bibnamefont {Fedrizzi}},\ and\ \bibinfo
  {author} {\bibfnamefont {B.~D.}\ \bibnamefont {Gerardot}},\ }\bibfield
  {title} {\bibinfo {title} {{A bright source of telecom single photons based
  on quantum frequency conversion}},\ }\href
  {https://doi.org/10.1063/5.0045413} {\bibfield  {journal} {\bibinfo
  {journal} {{Applied Physics Letters}}\ }\textbf {\bibinfo {volume} {118}},\
  \bibinfo {pages} {174003} (\bibinfo {year} {2021})}\BibitemShut {NoStop}%
\bibitem [{\citenamefont {Zeuner}\ \emph {et~al.}(2021)\citenamefont {Zeuner},
  \citenamefont {J{\"o}ns}, \citenamefont {Schweickert}, \citenamefont
  {{Reuterski{\"o}ld Hedlund}}, \citenamefont {{Nu{\~n}ez Lobato}},
  \citenamefont {Lettner}, \citenamefont {Wang}, \citenamefont {Gyger},
  \citenamefont {Sch{\"o}ll}, \citenamefont {Steinhauer}, \citenamefont
  {Hammar},\ and\ \citenamefont {Zwiller}}]{Zeuner.2021}%
  \BibitemOpen
  \bibfield  {author} {\bibinfo {author} {\bibfnamefont {K.~D.}\ \bibnamefont
  {Zeuner}}, \bibinfo {author} {\bibfnamefont {K.~D.}\ \bibnamefont
  {J{\"o}ns}}, \bibinfo {author} {\bibfnamefont {L.}~\bibnamefont
  {Schweickert}}, \bibinfo {author} {\bibfnamefont {C.}~\bibnamefont
  {{Reuterski{\"o}ld Hedlund}}}, \bibinfo {author} {\bibfnamefont
  {C.}~\bibnamefont {{Nu{\~n}ez Lobato}}}, \bibinfo {author} {\bibfnamefont
  {T.}~\bibnamefont {Lettner}}, \bibinfo {author} {\bibfnamefont
  {K.}~\bibnamefont {Wang}}, \bibinfo {author} {\bibfnamefont {S.}~\bibnamefont
  {Gyger}}, \bibinfo {author} {\bibfnamefont {E.}~\bibnamefont {Sch{\"o}ll}},
  \bibinfo {author} {\bibfnamefont {S.}~\bibnamefont {Steinhauer}}, \bibinfo
  {author} {\bibfnamefont {M.}~\bibnamefont {Hammar}},\ and\ \bibinfo {author}
  {\bibfnamefont {V.}~\bibnamefont {Zwiller}},\ }\bibfield  {title} {\bibinfo
  {title} {{On-Demand Generation of Entangled Photon Pairs in the Telecom
  C-Band with InAs Quantum Dots}},\ }\href
  {https://doi.org/10.1021/acsphotonics.1c00504} {\bibfield  {journal}
  {\bibinfo  {journal} {{ACS Photonics}}\ }\textbf {\bibinfo {volume} {8}},\
  \bibinfo {pages} {2337} (\bibinfo {year} {2021})}\BibitemShut {NoStop}%
\bibitem [{\citenamefont {{Da Lio}}\ \emph {et~al.}(2022)\citenamefont {{Da
  Lio}}, \citenamefont {Faurby}, \citenamefont {Zhou}, \citenamefont {Chan},
  \citenamefont {Uppu}, \citenamefont {Thyrrestrup}, \citenamefont {Scholz},
  \citenamefont {Wieck}, \citenamefont {Ludwig}, \citenamefont {Lodahl},\ and\
  \citenamefont {Midolo}}]{DaLio.2022b}%
  \BibitemOpen
  \bibfield  {author} {\bibinfo {author} {\bibfnamefont {B.}~\bibnamefont {{Da
  Lio}}}, \bibinfo {author} {\bibfnamefont {C.}~\bibnamefont {Faurby}},
  \bibinfo {author} {\bibfnamefont {X.}~\bibnamefont {Zhou}}, \bibinfo {author}
  {\bibfnamefont {M.~L.}\ \bibnamefont {Chan}}, \bibinfo {author}
  {\bibfnamefont {R.}~\bibnamefont {Uppu}}, \bibinfo {author} {\bibfnamefont
  {H.}~\bibnamefont {Thyrrestrup}}, \bibinfo {author} {\bibfnamefont
  {S.}~\bibnamefont {Scholz}}, \bibinfo {author} {\bibfnamefont {A.~D.}\
  \bibnamefont {Wieck}}, \bibinfo {author} {\bibfnamefont {A.}~\bibnamefont
  {Ludwig}}, \bibinfo {author} {\bibfnamefont {P.}~\bibnamefont {Lodahl}},\
  and\ \bibinfo {author} {\bibfnamefont {L.}~\bibnamefont {Midolo}},\
  }\bibfield  {title} {\bibinfo {title} {{A Pure and Indistinguishable
  Single--Photon Source at Telecommunication Wavelength}},\ }\href
  {https://doi.org/10.1002/qute.202200006} {\bibfield  {journal} {\bibinfo
  {journal} {{Advanced Quantum Technologies}}\ }\textbf {\bibinfo {volume}
  {5}},\ \bibinfo {pages} {2200006} (\bibinfo {year} {2022})}\BibitemShut
  {NoStop}%
\end{thebibliography}%
